%% file: main.tex
\documentclass[sigconf]{acmart}

\input{sections/setup}

\begin{document}

\title{Learning-based Models for Vulnerability Detection: An Extensive Study}

\author{Chao Ni}
\authornote{Chao Ni is the corresponding author.}
% \authornotemark[2]
% \orcid{0000-0002-2906-0598}`
\affiliation{%
  \institution{School of Software Technology, Zhejiang University}
  \city{Hangzhou}
  \state{Zhejiang}
  \country{China}
}
% \affiliation{%
%   \institution{Hangzhou High-Tech Zone (Binjiang) Blockchain and Data Security Research Institute}
%   % \streetaddress{8600 Datapoint Drive}
%   \city{Hangzhou}
%   % \state{Texas}
%   \country{China}
%   % \postcode{78229}
%   }
\email{chaoni@zju.edu.cn}

\author{Liyu Shen}
% \authornotemark[2]
\affiliation{%
  \institution{School of Software Technology, Zhejiang University}
  \city{Hangzhou}
  \country{China}}
\email{liyushen@zju.edu.cn}

\author{Xiaodan Xu}
\affiliation{%
  \institution{Zhejiang University}
  % \streetaddress{8600 Datapoint Drive}
  \city{Hangzhou}
  % \state{Texas}
  \country{China}
  % \postcode{78229}
  }
\email{xiaodanxu@zju.edu.cn}

\author{Xin Yin}
% \authornotemark[2] 
\affiliation{%
  \institution{School of Software Technology, Zhejiang University}
  \city{Hangzhou}
  \country{China}}
\email{xyin@zju.edu.cn}

\author{Shaohua Wang}
% \authornotemark[2]
\affiliation{%
  \institution{Central University of Finance and Economics}
  \city{Beijing}
  \country{China}}
\email{davidshwang@ieee.org}

% \settopmatter{printacmref=false,printccs=false}
% \renewcommand\footnotetextcopyrightpermission[1]{}
% \renewcommand{\shortauthors}{Chao and Wei, et al.}

\begin{abstract}
Though many deep learning-based models have made great progress in vulnerability detection, we have no good understanding of these models, which limits the further advancement of model capability, understanding of the mechanism of model detection, and efficiency and safety of practical application of models.
In this paper, we extensively and comprehensively investigate two types of state-of-the-art learning-based approaches (sequence-based and graph-based)  by conducting experiments on a recently built large-scale dataset.
We investigate seven research questions from five dimensions, namely \textit{model capabilities}, \textit{model interpretation}, \textit{model stability}, \textit{ease of use of model}, and \textit{model economy}.
We experimentally demonstrate the priority of sequence-based models and the limited abilities of both LLM (ChatGPT) and graph-based models.
We explore the types of vulnerability that learning-based models skilled in and reveal the instability of the models though the input is subtlely semantical-equivalently changed.
We empirically explain what the models have learned.
We summarize the pre-processing as well as requirements for easily using the models.
Finally, we initially induce the vital information for economically and safely practical usage of these models.
\end{abstract}

    


\maketitle

\input{sections/intro}
\label{sec:intro}

\input{sections/experiment}
\label{sec:experiment}

\input{sections/results}
\label{sec:results}

\input{sections/threats}
\label{sec:threats}

\input{sections/related_work}
\label{sec:related_work}

\input{sections/conclusion}

\label{sec:conclusion}

% \section*{Acknowledgements}{
% This research is supported by the National Natural Science Foundation of China (No. 62202419), the Fundamental Research Funds for the Central Universities (No. 226-2022-00064), the Ningbo Natural Science Foundation (No. 2022J184), and the State Street Zhejiang University Technology Center.
% }

% \begin{acks}
%     To Robert, for the bagels and explaining CMYK and color spaces.
% \end{acks}
% \section*{Acknowledgements}{
% This work was supported by the National Natural Science Foundation of China (Grant No.62202419 and No. 62172214), the Natural Science Foundation of Jiangsu Province, China (Grant No. BK20210279), the Key Research and Development Program of Zhejiang Province (No.2021C01105), and the Open Project Program of the State Key Laboratory of Mathematical Engineering and Advanced Computing (No. 2020A06).}

\balance
\bibliographystyle{ACM-Reference-Format}
\bibliography{main}

\end{document}

%% file: sections/setup.tex
\usepackage{svg}
\usepackage[ruled,vlined,lined,commentsnumbered]{algorithm2e}
\usepackage{amsmath,amsfonts}
\usepackage{array}
\usepackage{booktabs}
\usepackage[skip=1pt,labelfont=bf]{caption}
 \usepackage{calligra}
\usepackage{color, colortbl}
\usepackage{courier}
\usepackage{csvsimple}
\usepackage{enumitem}
\usepackage{fancybox}
\usepackage{fontenc}
\usepackage{subfigure}
\usepackage{graphicx}
\usepackage{listings}
\usepackage{longtable}
\usepackage{lscape}
\usepackage{makecell}
\usepackage{marvosym}
\usepackage{moreverb}
\usepackage{multicol}
\usepackage{multirow}
\usepackage{pifont}% 
\usepackage{newtxmath}
\usepackage{rotating}
\usepackage{setspace}
\usepackage[most]{tcolorbox}
\usepackage{threeparttable}
\usepackage{tikz}
\usepackage[normalem]{ulem}
\usepackage{url}
\usepackage{soul}
\usepackage{wasysym}
\usepackage{indentfirst}
\usepackage{textcomp}
\usepackage{xcolor}
\usepackage{wrapfig}
\usepackage{pdflscape}
\usepackage{hyperref}

\newcommand{\tabincell}[2]{\begin{tabular}{@{}#1@{}}#2\end{tabular}}
\usepackage{array}
\usepackage{balance}

\usepackage{microtype}

\newcolumntype{L}[1]{>{\raggedright\let\newline\\\arraybackslash\hspace{0pt}}m{#1}}
\newcolumntype{C}[1]{>{\centering\let\newline\\\arraybackslash\hspace{0pt}}m{#1}}
\newcolumntype{R}[1]{>{\raggedleft\let\newline\\\arraybackslash\hspace{0pt}}m{#1}}

\newboolean{showcomments}
\setboolean{showcomments}{true}
\ifthenelse{\boolean{showcomments}}
 { \newcommand{\mynote}[2]{
      \fbox{\bfseries\sffamily\scriptsize#1}
        {\small$\blacktriangleright$\textsf{\emph{#2}}$\blacktriangleleft$}}}
        { \newcommand{\mynote}[2]{}}

\usepackage{tikz}
\newcommand*\circled[1]{\tikz[baseline=(char.base)]{
            \node[shape=circle,draw,inner sep=1pt] (char) {#1};}}

\newcommand*\smallcircled[1]{\tikz[baseline=(char.base)]{
            \node[shape=circle,draw,inner sep=0.3pt] (char) {#1};}}

\newcommand{\datasetname}{MegaVul\xspace}

\newcommand{\intuition}[1]{
\begin{tcolorbox}[colback=white,boxrule=1pt,top=0pt,bottom=0pt,left=1pt,right=2pt,top=2pt,bottom=2pt]%[tile,size=fbox,boxsep=2mm,boxrule=0pt,top=0pt,bottom=0pt,borderline={0.5mm}{0pt}{black!70!white},colback=black!5!white]
\em #1
\end{tcolorbox}
}

\AtBeginDocument{%
  \providecommand\BibTeX{{%
    \normalfont B\kern-0.5em{\scshape i\kern-0.25em b}\kern-0.8em\TeX}}}

\setcopyright{acmcopyright}
\copyrightyear{2024}
\acmYear{2024}
\acmDOI{XXXXXXX.XXXXXXX}

%% file: sections/intro.tex
\section{Introduction}

Automated vulnerability detection is a fundamental problem in system security and learning-based vulnerability detection approaches have achieved promising results in recent years.
Notably, several research has demonstrated that deep learning (DL)-based approaches can achieve both high accuracy (up to 95\%)~\cite{li2018vuldeepecker,li2021sysevr,russell2018automated,li2017large,maiorca2019digital,suarez2017droidsieve,zhou2019devign}  and high F1-score (up to 90\%)~\cite{fu2022linevul,song2022hgvul} at detecting vulnerabilities.

Despite the remarkable success of DL models at detecting vulnerabilities, we seem to know little about the models themselves.
For example, whether they can be used effectively and reliably in detecting real-world or most dangerous vulnerabilities, what kind of vulnerability the models are skilled in detecting, what kind of features these models have learned, whether the models can stably perform on semantically equal functions, what is the complexity and what is the cost if we want to use DL models, whether it will damage our privacy.
Figuring out the answers to these questions can help us better develop and apply the models in practical usage, especially in the era of large language models.

Some work~\cite{steenhoek2023empirical,chakraborty2021deep} also aim at investigating the characteristics or capabilities of deep learning-based models, but there still are a few limitations in guiding practice use.
For example, 
(1) The conclusions obtained by different works may be inconsistent. {Some works~\cite{wang2023deepvd,Wen2023vuldetect} conclude that graph-based DL models outperform sequence-based models, while some works~\cite{fu2022linevul,ni2023distinguishing} achieve the opposite observation.}
(2) Existing studies only include graph-based or sequence-based models, but the current eye-catching large language model (e.g., ChatGPT~\cite{openaichatgpt}) has not been comprehensively studied, which may cause the conclusion to be biased.
(3) Previous studies only consider the capability and explanation of DL models but do not consider the impacts on users: ease of use of the model and cost of using the model.

In this paper, we systematically and comprehensively investigate the characteristics of several SOTA deep learning-based vulnerability
detection~\cite{zhou2019devign,chakraborty2021deep,li2021vulnerability,fu2022linevul,ni2023distinguishing} and construct research questions to understand these models with the goal of distilling lessons and guidelines for better practical usage.
We primarily focus on the graph-based and sequence-based vulnerability detection models at function granularity. 
To the best of our knowledge, this is the first paper that systematically investigates SOTA learning-based models across a wide range of aspects in the era of LLMs.

\begin{table*}[!th]
% \vspace{-0.3cm}
    \centering
    \caption{Insights and Takeaways: Evaluation on Extensive Newly Built Dataset (\datasetname)}
    
    \resizebox{\linewidth}{!}
    {
      \begin{tabular}{l|l}
      \toprule
      \textbf{Dimension} &{\textbf{Findings or Insights}}\\
      \midrule
      \textbf{Capability} & 

      \tabincell{l}{
      
      \textbf{\circled{1}}. Sequence-based models outperform graph-based models. \\
      \textbf{\circled{2}}. ChatGPT is not yet competent for vulnerability detection and different prompts enable varying ability.\\
      \textbf{\circled{3}}. Different models have their own advantages in detecting different types of vulnerabilities. \\
\textbf{\circled{4}}. Sequence-based models are skilled in ``Input Validation''. \\
\textbf{\circled{5}}. Graph-based models are skilled in ``API Abuse'', ``Input Validation'' and ``Security Features''. \\
\textbf{\circled{6}}. Sequence-based models especially SVulD its promising potential in practical usage.
} \\
\midrule
       \textbf{Interpretation} & \tabincell{l}{
       \textbf{\circled{7}}. Both graph-based and sequence-based methods  focus on two types of statements: ``Function Calls'' and ``Field Expressions''.\\
       \textbf{\circled{8}}. Learning-based models still have a limited ability to distinguish vulnerable functions from non-vulnerable functions.\\
       \textbf{\circled{9}}.  Feeding external called function information sequence-based method could further improve sequence-based models'  ability. } \\
       \midrule
      \textbf{Stability}& \tabincell{l}{
       \textbf{\smallcircled{10}}. All the learning-based models are unstable to input changes even if these changes are semantically equivalent.\\
       \textbf{\smallcircled{11}}. Sequence-based models perform more stably than graph-based models.
       }\\
       \midrule
       \textbf{Ease of Use}& \tabincell{l}{
\textbf{\smallcircled{12}}.
Sequence-based models: easy to deploy, limited input size, fine-tuning, open-source, privacy-safe. 
\\ChatGPT: easy to use, larger input size; privacy-unsafe.\\
\textbf{\smallcircled{13}}.  Graph-based models: complete function, complex configurations, limited input size, fine-tuning, open-source, privacy-safe.
       }\\
       \midrule
        \textbf{Economy}& \tabincell{l}{
        \textbf{\smallcircled{14}}. Graph-based models need large amounts of time for data preprocessing, but they typically train and infer fast.\\ 
        Sequence-based models do not involve data preprocessing, with a comparable training time and longer inference time.\\
        \textbf{\smallcircled{15}}. ChatGPT is the most economical solution.
        }  \\
         \bottomrule
       \end{tabular}
       }
    %    \vspace{-0.3cm}
       \label{tab:takeaway}
\end{table*}

In particular, we conduct seven research questions and classify them into the following dimensions: \textbf{D1: model capabilities}, \textbf{D2: model interpretation}, \textbf{D3: model stability}, \textbf{D4: ease of use of model} and \textbf{D5: model economy}.  
More precisely, our first goal is to understand the capabilities of the learning models for vulnerability detection tasks, especially aiming at asking the following research questions:
\begin{itemize}
    \item \textbf{RQ-1}: How do learning-based approaches perform on vulnerability detection? What are the variabilities across different models?
     \item \textbf{RQ-2}: What types of vulnerabilities are learning-based approaches skilled in detecting?
     \item \textbf{RQ-3}: Are Large Language Models capable of detecting vulnerabilities?
\end{itemize}

Our second study aims at the model interpretation. 
We adopt the state-of-the-art explanation tools to investigate what the model has learned as follows:
\begin{itemize}
    \item  \textbf{RQ-4}: What source code information does the learning-based model focus on? Do different types of learning-based models agree on similar important code features?
\end{itemize}

Our third study targets the stability of the studied learning-based models by investigating the impacts of semantically equivalent subtle modifications to their input.
\begin{itemize}
    \item \textbf{RQ-5}: Do learning-based models agree on the vulnerability detection results with themselves when the input is insignificantly changed?
\end{itemize}

Our fourth study focuses on the ease of use by investigating the various efforts to build an effective model.
\begin{itemize}
    \item \textbf{RQ-6}: What types of efforts should be paid before using a model? In what scenarios can learning-based models be applied?
\end{itemize}

Finally, our study focuses on the economy.
We want to investigate the cost when adopting the models to detect vulnerability.
\begin{itemize}
    \item \textbf{RQ-7}: What are the costs caused by models from both time and economic aspects?
\end{itemize}

To answer the aforementioned research questions, we investigate two types of state-of-the-art learning-based models.
These models used different deep learning architectures (e.g., transformer~\cite{vaswani2017attention} or graph~\cite{zhou2020graph}).
Besides, to extensively and comprehensively analyze the models' ability, we conduct experiments on the recently built dataset named MageVul~\cite{ni2024megavul}, which contains real-world projects' vulnerabilities by crawling more newly discovered vulnerabilities.
Then, we carefully design experiments to discover the findings by answering seven RQs.
Eventually, the main contribution of our work is summarized as follows and takeaway findings are shown in Table~\ref{tab:takeaway}.
\begin{itemize}
    % \item We construct an extensive real-world vulnerability dataset for vulnerability analysis.
    \item We conduct an extensive comparison among learning-based approaches on vulnerability detection including ChatGPT.
    \item We design seven RQs grouped into five important dimensions to understand learning-based approaches comprehensively.
    \item {We release our reproduction package for further study}~\cite{replication}.
\end{itemize}

%% file: sections/experiment.tex
\section{Experimental Setup}

% \cn
{
In this section, we first introduce our studied dataset.
Following that, we briefly describe the studied learning-based models and evaluation metrics.
Finally, the implementation details are presented.}

% \vspace{-0.3cm}
\input{sections/dataset}

\label{sec:dataset}

% \vspace{-0.9cm} 
\subsection{Studied Baselines and Evaluation Metrics}

\textit{\ul{Baselines}}. To comprehensively compare the performance of existing work, in this paper, we consider the five state-of-the-art learning-based software vulnerability detection approaches and these approaches can be further divided into two finer categories: graph-based ones and sequence-based ones.
The former group contains three methods (i.e.,  Devign~\cite{zhou2019devign}, {\sc ReVeal}~\cite{chakraborty2021deep} and  {\sc IVDetect}~\cite{li2021vulnerability}) and they transform source code into a graph to complexly represent its semantic.
The latter group contains two approaches (i.e., LineVul~\cite{fu2022linevul} and SVulD~\cite{ni2023distinguishing}) and they treat the source code as the sequence of tokens to simply represent its semantic.
Here, we briefly introduce these methods to make our paper self-contained.

\textit{\ul{Evaluation Metrics}}.
To comprehensively investigate the performance of learning-based models for vulnerability detection, we adopt the following widely used evaluation metrics~\cite{ni2022best,ni2023distinguishing,zhou2019devign,fu2022linevul}: Accuracy(A), Precision(P), Recall(R), and F1-score(F1).

\begin{table}[htbp]
% \vspace{-0.3cm}
  \centering
  \caption{The statistics of \datasetname (C/C++)}
    \resizebox{\linewidth}{!}
    {
    \begin{tabular}{l|r}
    \toprule
 \textbf{Attributes} & \textbf{\datasetname}\\
\midrule

Number of Projects & 736 \\
Number of CVE IDs & 5,714 \\
Date range of crawled CVEs & \tabincell{r}{2013/01$\sim$2023/04 \\(continuously updating)}\\
Number of CWE IDs & 159 \\
Number of Commits & 6,437 \\
Number of Crawled Code Hosting Platforms &  17 \\
Number of Vul/Non-Vul Function & 14,216/377,185 \\ 
Function Extract Strategy  & Tree-sitter \\ 
Dimensions of Information & 
% \makecell{CVSS,Diff info, Commit structure \\ \textit{Nine-granularity} abstraction \\ \textit{Joern} graph,function signature}
% & CVSS,Diff info \\
6\\
Code Availability & Full  \\
\bottomrule
\end{tabular}}
% \vspace{-0.3cm}
\label{tab:vul4c}
\end{table}

% \vspace{-0.3cm}
\subsection{Implementation Details}

% \textcolor{blue}
{\textit{\ul{Data Splitting}}. Similar to existing work~\cite{fu2022linevul,li2021vulnerability}, we adopt the same data splitting approach: 80\%:10\%:10\%.
More precisely, the whole dataset is split into 80\% of training
data, 10\% of validation data, and 10\% of testing data.
Meanwhile, we also keep the class distribution as same as the original ones  in training data, validation data, and testing data.
}

{
\textit{\ul{Model Implementation}}. 
Regarding ReVeal, IVDetect, LineVul, and SVulD, we utilize their publicly available source code and perform fine-tuning with the default parameters provided in their original code. 
Considering Devign's code is not publicly available, we make every effort to replicate its functionality and achieve similar results on the original paper's dataset.
All these models are implemented using the PyTorch~\cite{pytorch} framework by fully adopting the pre-trained models hosted on  Huggingface~\cite{huggingface}. 
Additionally, we incorporate interpretability into all studied models.
The fine-tuning process is performed on NVIDIA RTX 3090 graphics card.
For the LLMs, we use the state-of-the-art ChatGPT~\cite{openai2022chatgpt} model and set the number of few-shot learning examples between 1 and 6 to fill the context window (i.e. 4,096 tokens).
We also instruct ChatGPT to output the results in JSON format to  facilitate the automatic organization of the data.
}

%% file: sections/dataset.tex
\subsection{Studied Dataset}

Though many vulnerability-related datasets have been proposed, there are still some limitations that impact the verification of proposed models, including (1) \textit{unreal vulnerability} (i.e., SARD~\cite{sard} is artificially synthesized), (2) \textit{unreal data distribution} (i.e., balanced distribution in Devign~\cite{zhou2019devign}), (3) \textit{limited diversity} (i.e., limited projects and vulnerability types in ReVeal~\cite{chakraborty2021deep}), (4) \textit{limited newly disclosed vulnerabilities} (i.e., no updated to Big-Vul~\cite{fan2020ac} covering the period only from 2003 to 2019), and (5) \textit{low-quality of dataset} (i.e., incomplete function, erroneously merged functions, missed commit message in Big-Vul~\cite{fan2020ac}).

To address the issues above, recently, Ni et al.~\cite{ni2024megavul} built a large-scale, high-quality, data-rich, multi-dimensional C/C++ and Java dataset named \datasetname by crawling data from more open-source repositories, adopting sophisticated filtering strategies to improve the quality, and employing advanced techniques to extract complete functions.
In addition to collecting the raw functions, MegaVul also provides more dimension information on the function, including the nine types of granularity abstraction of functions (i.e., \textit{FUNC}, \textit{VAR}, \textit{STRING}, etc.), various types of function representations (i.e., \textit{AST}, \textit{PDG}),
and the details of function modifications (i.e., \textit{diff}).
In summary, MageVul collects 17 Git-based code hosting platforms from 349 websites that had referenced the CVEs more than 100 times. 
The web-based code hosting platforms can be categorized into five main categories: GitHub, GitLab, GitWeb, CGit, and Gitiles.
Considering both the regular updates (i.e., update every six months) and the stability of MegaVul, we consider the C/C++ dataset released in October 2023 for experiments.
That is, for the C/C++ version,  MegaVul contains 8,334 commits from 198,994 CVEs.
Table~\ref{tab:vul4c} presents the statistical information of MegaVul and more details can be referred to their original work~\cite{ni2024megavul}.

%% file: sections/results.tex
% \vspace{-0.2cm}
\section{Research Question and Findings}

% \cn
{
In this section, we divide our seven research questions into five dimensions: \textit{D1: model capabilities}, \textit{D2: model interpretation}, \textit{D3: model stability}, \textit{ease of use of model} and \textit{model economy}.
For each RQ, we introduce the objective, the experimental setup, the results, and our findings.
}

\subsection{D1: Capabilities of Learning-based Models for Vulnerability Detection}

\noindent
{$\bullet$\bf [RQ-1]: \ul{How do learning-based approaches perform on vulnerability detection? What are the variabilities across different models?}}
\label{sec:rq1}

\noindent
\textbf{Objective}.
% \cn
% {Benefiting from the powerful representation capability of deep neural networks, 
Many deep learning-based vulnerability detection approaches have been proposed~\cite{zhou2019devign,li2021vulnerability,ni2023distinguishing,hanif2022vulberta,li2018vuldeepecker,cao2022mvd,wang2023deepvd} and they mainly focus on function-level vulnerability detection, treating the source code in different ways.
That is, some approaches~\cite{zhou2019devign,li2021vulnerability} consider the complex structure inside a function and transform it into a graph, while some approaches~\cite{fu2022linevul,ni2023distinguishing} simply treat it as a sequence of tokens without considering its structure (i.e., sequence-based).
Though these methods have been well compared in previous
studies~\cite{steenhoek2023empirical,ni2023distinguishing,wang2023deepvd}, their experiments are usually conducted on a limited or small-scale dataset, which may impact the consistency of models' capabilities.
For example, Wen et al.~\cite{Wen2023vuldetect} concluded that a complex graph-based model embedding a function by considering the program structure can yield better performance than sequence-based models.
However, according to recent works~\cite{ni2023distinguishing,fu2022linevul}, sequence-based models seem to outperform graph-based ones.
Meanwhile, recently large language models (especially ChatGPT) have attracted much attention since their powerful ability can be easily adapted to various types of downstream tasks, including vulnerability detection.
However, there are no comparisons between ChatGPT and existing models.
Considering these issues, we want to conduct an extensive study to comprehensively compare learning-based models' abilities.

\noindent
\textbf{Experimental Setup}.
% \cn
{We consider three graph-based approaches (i.e., Devign~\cite{zhou2019devign}, {\sc ReVeal}~\cite{chakraborty2021deep} and {\sc IVDetect}~\cite{li2021vulnerability}) and two sequence-based approaches (i.e., LineVul~\cite{fu2022linevul} and SVulD~\cite{ni2023distinguishing}).
Meanwhile, to comprehensively compare the performance difference, we adopt the currently largest dataset MageVul.
Since graph-based approaches usually need to obtain the structure information of the function (e.g., CFG, DFG), we adopt the same toolkit with Joern to transform functions and drop the fail-passed cases.
Finally, the filtered dataset ({391,401}, shown in Table~\ref{tab:vul4c}) is used for evaluation and we follow previous work ~\cite{fu2022linevul, ni2022defect} to split the dataset into the training data (i.e., 80\%), validating data (i.e., 10\%), and testing data (i.e., 10\%). 
We also keep the distribution as same as the original ones in training, validating, and testing data.
}

% \cn
{
Furthermore, we also consider another LLM model ChatGPT and it also treats source code as a sequence of tokens.
We prompt ChatGPT with an in-context learning setting and equip it with {1$\sim$6} examples selected from the same projects (i.e., cf. Section~\ref{sec:rq3} for details).
ChatGPT is a commercial conversation-based LLM model developed by OpenAI and can only be accessed by its API or web interface.
Considering the large-scale testing size (i.e., {38,749 functions}) as well as the substantial cost when interacting with ChatGPT, {we follow previous work~\cite{croft2023data}} to statistically sample some cases with 95\% confidence and we conduct experiments on these sampled functions.
}

\begin{table}[htbp]
  % \vspace{-0.3cm}
  \centering
  \caption{The comparisons among learning-based approaches}
  \resizebox{\linewidth}{!}
  {
  \begin{threeparttable}
    \begin{tabular}{llcccc}
    \toprule
    \textbf{Types} & \textbf{Models} & \textbf{Accuracy} & \textbf{Recall} & \textbf{Precision} & \textbf{F1} \\
    \midrule
    \multirow{3}[2]{*}{\textbf{\tabincell{l}{Graph\\Based}}} & Devign & 0.742  & 0.622  & 0.068  & 0.122  \\
          & Reveal & 0.780  & 0.545  & 0.070  & 0.125  \\
          & IVDetect & 0.792  & 0.582  & 0.080  & 0.141  \\
    \midrule
    \multirow{3}[2]{*}{\textbf{\tabincell{l}{Sequence\\Based}}} & LineVul & \textbf{0.962} & 0.593  & \textbf{0.117} & \textbf{0.195} \\
          & SVulD & 0.822  & \textbf{0.637} & 0.100  & 0.172  \\
          \cmidrule{2-6}
          &\cellcolor{lightgray}ChatGPT$^\ast$& \cellcolor{lightgray}0.932 $\pm$0.015  & \cellcolor{lightgray}0.125 $\pm$0.020 & \cellcolor{lightgray} 0.057 $\pm$0.014& \cellcolor{lightgray}0.078 $\pm$0.016 \\
    \bottomrule
    \end{tabular}%
    $^\ast$ Notice that the performance of ChatGPT is calculated on statistical sampling with 95\% confidence.
    \end{threeparttable}
  
    }
  \label{tab:rq1}%
    % \vspace{-0.3cm} 
\end{table}%

\noindent
\textbf{Results}.
% \cn
{Table~\ref{tab:rq1} shows the comparison results and the best ones are highlighted in bold.
From the results, we can draw the following observations:
(1) Surprisingly, the sequence-based modes perform better than the graph-based models in terms of all evaluated metrics, which indicates that we may not be concerned about the complex code structure when utilizing deep learning techniques to build a vulnerability detector.
(2) Among sequence-based models, these methods have a complementary ability to detect vulnerabilities.
More precisely, LineVul performs better in terms of \textit{Accuracy}, \textit{Precision}, and \textit{F1-score}, while SVulD performs better in terms of \textit{Recall}.
It means that sequence-based models can be used for different usage scenarios, for example, LineVul for high \textit{Precision} and SVulD for high \textit{Recall}.
(3) Though ChatGPT's performance is obtained on the statistically sampled dataset with 95\% confidence, its performance is still far away from the existing SOTA baselines, especially in terms of \textit{Recall}, \textit{Accuracy} and \textit{F1-score}, which means that currently, ChatGPT is not yet competent for vulnerability detection tasks.
(4) Considering the original goal difference between ChatGPT (i.e., target various tasks including QA, NLP, SE, etc.) and existing sequence models (i.e., LineVul and SVulD, target exclusively vulnerability detection), we find that it is necessary to build a vulnerability detection targeted model to make further progress in the field.
}

% \vspace{-0.2cm}
\intuition{
\textbf{Finding 1}: (1) Sequence-based vulnerability detection models achieve better performance than graph-based models.
(2) ChatGPT is not yet competent  for software vulnerability detection and it is necessary to build a vulnerability detection targeted sequence model.
}
% \vspace{-0.2cm}

\noindent
{$\bullet$ \bf [RQ-2]: \ul{What types of vulnerabilities are learning-based approaches skilled in detecting?}}

\noindent
\textbf{Objective}.
% \cn
{
Many types of models have been proposed for vulnerability detection and among them, graph-based and sequence-based are the promising ones.
However, different approaches may have their own advantages in detecting different types of vulnerabilities.
Figuring out their expertise can better guide us in practical usage.
Therefore, we want to analyze what are the types of vulnerabilities that each learning-based approach skilled in.
}

\noindent
\textbf{Experimental Setup}.
% \cn
{
We make an analysis of each approach's  performance on vulnerability types in the testing dataset and pick up the Top-10 vulnerability types that are most correctly classified for each method.
Besides, following previous work~\cite{tsipenyuk2005seven}, we can group  the vulnerabilities into 7 categories, namely 
\textit{Input Validation and Representation}, \textit{API Abuse}, \textit{Security Features}, \textit{Time and State}, \textit{Errors}, \textit{Code Quality}, and \textit{Encapsulation}, shown under vulnerability types in Table~\ref{tab:seven_kingdom_types}.
Specifically, ``Input Validation and Representation'' is caused by
metacharacters, alternate encodings, and numeric representations.
% Security problems result from trusting input. 
e.g.,  CWE-787 ``Out-of-bounds Write''.
% The issues include: Buffer Overflows, Cross-Site Scripting attacks, SQL Injection, and many others
The mapping of the complete CWE list to these groups can be found in our dataset.
``API Abuse'' is commonly caused by the caller failing to honor the end of a contract between the caller and callee, e.g., CWE-252 ``Unchecked Return Value''.
``Security Features'' mainly concerns topics like authentication, access control, confidentiality, cryptography, and privilege management, e.g., CWE 359 ``Privacy Violation''.
``Time and State'' mainly concerns the time and state in distributed computation for more than one component to communicate correctly by sharing the state and time, e.g., CWE-833 ``Deadlock''.
``Errors'' relates to a class of API that handles errors, e.g., CWE-1069 ``Empty Exception Block''. 
``Code Quality'' mainly concerns the unpredictable behavior caused by poor code quality.
It leads to poor usability for a user and  provides an opportunity to stress the system in unexpected ways for an attacker, e.g., CWE-476 ``NULL Pointer Dereference''.
``Encapsulation'' aims to draw strong boundaries of operations, e.g., CWE-501 ``Trust Boundary Violation''.
Notice that there are no instances in \textit{Encapsulation} and no method can correctly detect vulnerability belonging to both ``Errors'' and ``Time and State'', we do not need to analyze them further.
}

% Table generated by Excel2LaTeX from sheet 'rq2'
\begin{table}[htbp]
  % \vspace{-0.3cm}
  \centering
  \caption{Seven Types of Vulnerability}
  \resizebox{\linewidth}{!}
  {
  \begin{threeparttable}
    \begin{tabular}{lrrl}
    \toprule
\textbf{Vulnerability Type} & \textbf{\# Total} & \textbf{\# Testing} &\textbf{CWE Example}\\
\midrule
Input Validation and Representation & 2,887 & 294 &  CWE-20\\
Code Quality & 1,543 & 170 & CWE-416\\
Security Features & 376 & 32 & CWE-284 \\
API Abuse & 17 & 4 & CWE-252 \\
Time and State & 13 & 1 & CWE-367 \\
Errors &  7 & 1 & CWE-388 \\
Encapsulation & - & - & CWE-501\\ 
\bottomrule
\end{tabular}
$^\ast$No C/C++ instance in MegaVul belongs to ``Encapsulation''.
\end{threeparttable}
  
}
\label{tab:seven_kingdom_types}
  % \vspace{-0.3cm}
\end{table}

% \begin{figure}[htbp]
%     \centering
%     \includesvg[width=0.7\linewidth]{figs/kingdons.svg}
%     \caption{Performance on Vulnerability Type}
%     \label{fig:vul_types}
% \end{figure}

% \cn
{Besides, we also analyze the Top-25 
Most Dangerous Software Weaknesses\footnote{https://cwe.mitre.org/top25/archive/2023/2023\_top25\_list.html}  to figure out the promising approach in detecting the most dangerous vulnerabilities in practice. 
Notice that six of the most dangerous CWEs (i.e., CWE-352, CWE-434, CWE-502, CWE-77, CWE-798, and CWE-306) are not included in the testing dataset, we, therefore, delete them from the list.
All the results are from the ones in RQ-1.
}

% Table generated by Excel2LaTeX from sheet 'rq2'
\begin{table}[htbp]
    % \vspace{-0.2cm}
  \centering
   \caption{Top-10 correctly detected CWE by each method}
  \resizebox{\linewidth}{!}
  {
    \begin{tabular}{llrrrr}
    \toprule
    \rowcolor{lightgray}\textbf{Approach} & \textbf{Top-1} & \textbf{Top-2} & \textbf{Top-3} & \textbf{Top-4} & \textbf{Top-5} \\
    \midrule
    \textbf{Devign} & CWE-78   [2/2] & CWE-918 [2/2] & CWE-276 [2/2] & CWE-863 [3/4] & CWE-94   [3/4] \\
    \textbf{Reveal} & CWE-94   [4/4] & CWE-79   [2/2] & CWE-918 [2/2] & CWE-89   [1/1] & CWE-287 [3/4] \\
    \textbf{IVDetect} & CWE-89   [1/1] & CWE-863 [3/4] & CWE-94   [3/4] & CWE-20   [61/86] & CWE-119 [104/148] \\
    \midrule
    \textbf{LineVul} & CWE-918 [2/2] & CWE-89   [1/1] & CWE-863 [3/4] & CWE-20   [63/86] & CWE-119 [107/148] \\
    \textbf{SVulD} & CWE-79   [2/2] & CWE-918 [2/2] & CWE-190 [29/35] & CWE-787 [65/79] & CWE-863 [3/4] \\
    \textbf{ChatGPT} & CWE-476 [1/3] & CWE-125 [1/3] & CWE-787 [1/5] & /     & / \\
    \midrule
    
    \rowcolor{lightgray}\textbf{Approach} & \textbf{Top-6} & \textbf{Top-7} & \textbf{Top-8} & \textbf{Top-9} & \textbf{Top-10} \\
    \midrule
    \textbf{Devign} & CWE-269 [3/4] & CWE-287 [3/4] & CWE-787 [58/79] & CWE-125 [47/70] & CWE-190 [23/35] \\
    \textbf{Reveal} & CWE-787 [56/79] & CWE-20   [53/86] & CWE-190 [21/35] & CWE-125 [41/70] & CWE-119 [85/148] \\
    \textbf{IVDetect} & CWE-787 [53/79] & CWE-190 [23/35] & CWE-125 [41/70] & CWE-476 [33/62] & CWE-362 [22/43] \\
    \midrule
    \textbf{LineVul} & CWE-787 [56/79] & CWE-190 [23/35] & CWE-125 [44/70] & CWE-22   [4/7] & CWE-362 [24/43] \\
    \textbf{SVulD} & CWE-287 [3/4] & CWE-119 [104/148] & CWE-20   [59/86] & CWE-362 [28/43] & CWE-125 [42/70] \\
    \textbf{ChatGPT} & /     & /     & /     & /     & / \\
    \bottomrule
    \end{tabular}%
    }
  \label{tab:rq2-1}%

\end{table}%

\begin{figure}[htbp]
  % \vspace{-0.3cm}
  \centering
  \includegraphics[width=\linewidth]{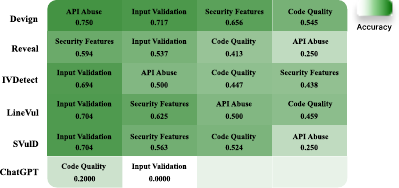}
  \caption{Performance on Vulnerability Type}
  \label{fig:vul_types_acc}
  % \vspace{-0.3cm}
\end{figure}

% \begin{figure}[htbp]
%     \vspace{-0.3cm}
%     \centering
%     \includesvg[width=.85\linewidth]{figs/kingdons_acc.svg}
%     \caption{Performance on Vulnerability Type}
%     \label{fig:vul_types_acc}
%     \vspace{-0.3cm}
% \end{figure}

% Table generated by Excel2LaTeX from sheet 'rq2'
\begin{table}[htbp]
  \centering
  \caption{The performance comparison among studied six approaches on Top-25 most risk CWE}
  \resizebox{\linewidth}{!}{
  \begin{threeparttable}
    \begin{tabular}{ccrrr|rrr}
    \toprule
    \multirow{2}[1]{*}{\textbf{ID}} & \multirow{2}[1]{*}{\textbf{CWE}} & \multicolumn{3}{c}{\textbf{Graph-based}} & \multicolumn{3}{c}{\textbf{Sequence-based}} \\
\cmidrule{3-8}          &       & \textbf{Devign} & \textbf{Reveal} & \textbf{IVdetect} & \textbf{LineVul} & \textbf{SVulD} & \textbf{ChatGPT$^\ast$} \\
    \midrule
    \multicolumn{1}{l}{1} & \multicolumn{1}{l}{CWE-787} & 53/79 & 41/79 & 53/79 & 56/79 & \cellcolor[rgb]{ .886,  .937,  .855}\textbf{67/79} & 1/5 \\
    \multicolumn{1}{l}{2} & \multicolumn{1}{l}{CWE-79} & \cellcolor[rgb]{ .886,  .937,  .855}\textbf{2/2} & 1/2   & 0/2   & 1/2   & 1/2   & 0/0 \\
    \multicolumn{1}{l}{3} & \multicolumn{1}{l}{CWE-89} & \cellcolor[rgb]{ .886,  .937,  .855}\textbf{1/1} & \cellcolor[rgb]{ .886,  .937,  .855}\textbf{1/1} & \cellcolor[rgb]{ .886,  .937,  .855}\textbf{1/1} & \cellcolor[rgb]{ .886,  .937,  .855}\textbf{1/1} & 0/1   & 0/0 \\
    \multicolumn{1}{l}{4} & \multicolumn{1}{l}{CWE-416} & 34/72 & 26/72 & 26/72 & 31/72 & \cellcolor[rgb]{ .886,  .937,  .855}\textbf{35/72} & 0/2 \\
    \multicolumn{1}{l}{5} & \multicolumn{1}{l}{CWE-78} & 0/2   & 0/2   & 0/2   & \cellcolor[rgb]{ .886,  .937,  .855}\textbf{1/2} & \cellcolor[rgb]{ .886,  .937,  .855}\textbf{1/2} & 0/0 \\
    \midrule
    \multicolumn{1}{l}{6} & \multicolumn{1}{l}{CWE-20} & 56/86 & 43/86 & 61/86 & \cellcolor[rgb]{ .886,  .937,  .855}\textbf{63/86} & 61/86 & 0/4 \\
    \multicolumn{1}{l}{7} & \multicolumn{1}{l}{CWE-125} & \cellcolor[rgb]{ .886,  .937,  .855}\textbf{47/70} & 41/70 & 41/70 & 44/70 & 41/70 & 1/3 \\
    \multicolumn{1}{l}{8} & \multicolumn{1}{l}{CWE-22} & \cellcolor[rgb]{ .886,  .937,  .855}\textbf{5/7} & 4/7   & 3/7   & 4/7   & 4/7   & 0/0 \\
    \multicolumn{1}{l}{11} & \multicolumn{1}{l}{CWE-862} & \cellcolor[rgb]{ .996,  .78,  .812}0/1 & \cellcolor[rgb]{ .996,  .78,  .812}0/1 & \cellcolor[rgb]{ .996,  .78,  .812}0/1 & \cellcolor[rgb]{ .996,  .78,  .812}0/1 & \cellcolor[rgb]{ .996,  .78,  .812}0/1 & 0/0 \\
    \multicolumn{1}{l}{12} & \multicolumn{1}{l}{CWE-476} & 34/62 & 27/62 & 33/62 & 30/62 & \cellcolor[rgb]{ .886,  .937,  .855}\textbf{37/62} & 1/3 \\
    \multicolumn{1}{l}{13} & \multicolumn{1}{l}{CWE-287} & 3/4   & \cellcolor[rgb]{ .886,  .937,  .855}\textbf{4/4} & 2/4   & 2/4   & 3/4   & 0/0 \\
    \multicolumn{1}{l}{14} & \multicolumn{1}{l}{CWE-190} & 27/35 & 23/35 & 23/35 & 23/35 & \cellcolor[rgb]{ .886,  .937,  .855}\textbf{29/35} & 0/1 \\
    \multicolumn{1}{l}{17} & \multicolumn{1}{l}{CWE-119} & 100/148 & 70/148 & 105/148 & \cellcolor[rgb]{ .886,  .937,  .855}\textbf{107/148} & 103/148 & 0/0 \\
    \multicolumn{1}{l}{19} & \multicolumn{1}{l}{CWE-918} & \cellcolor[rgb]{ .886,  .937,  .855}\textbf{2/2} & 0/2   & 1/2   & \cellcolor[rgb]{ .886,  .937,  .855}\textbf{2/2} & \cellcolor[rgb]{ .886,  .937,  .855}\textbf{2/2} & 0/0 \\
    \multicolumn{1}{l}{21} & \multicolumn{1}{l}{CWE-362} & \cellcolor[rgb]{ .886,  .937,  .855}\textbf{24/43} & 16/43 & 22/43 & \cellcolor[rgb]{ .886,  .937,  .855}\textbf{24/43} & \cellcolor[rgb]{ .886,  .937,  .855}\textbf{24/43} & 0/0 \\
    \multicolumn{1}{l}{22} & \multicolumn{1}{l}{CWE-269} & \cellcolor[rgb]{ .886,  .937,  .855}\textbf{3/4} & 2/4   & 2/4   & 1/4   & 2/4   & 0/0 \\
    \multicolumn{1}{l}{23} & \multicolumn{1}{l}{CWE-94} & \cellcolor[rgb]{ .886,  .937,  .855}\textbf{3/4} & 2/4   & 3/4   & 0/4   & 2/4   & 0/0 \\
    \multicolumn{1}{l}{24} & \multicolumn{1}{l}{CWE-863} & 1/4   & 1/4   & \cellcolor[rgb]{ .886,  .937,  .855}\textbf{3/4} & \cellcolor[rgb]{ .886,  .937,  .855}\textbf{3/4} & \cellcolor[rgb]{ .886,  .937,  .855}\textbf{3/4} & 0/0 \\
    \multicolumn{1}{l}{25} & \multicolumn{1}{l}{CWE-276} & \cellcolor[rgb]{ .886,  .937,  .855}\textbf{2/2} & 0/2   & 1/2   & 1/2   & 0/2   & 0/0 \\
    \midrule
    \multicolumn{2}{c}{\textbf{\# Wins (628)}} & \textbf{397} & \textbf{302} & \textbf{380} & \textbf{394} & \textbf{415} & 3/18 \\
    \bottomrule
    \end{tabular}%
         $^\ast$ Notice that the performance of ChatGPT is calculated on statistical sampling with 95\% confidence.
  \end{threeparttable}
    }
  % \vspace{-0.3cm}
  \label{tab:rq2-2}%
\end{table}%

\noindent
\textbf{Results}.
Table~\ref{tab:rq2-1} shows the results of the Top 10 CWE that each method performs well and Fig.~\ref{fig:vul_types_acc} shows the performance of different vulnerability groups.
From the results, we observe that:
(1) Each method performs variously on different vulnerability types which indicates their complementary ability.
(2) Overall, all methods perform best in ``Input Validation and Representation'' and perform relatively worst in ``API Abuse''.
(3) Sequence-based approaches perform similarly, but graph-based approaches perform differently.
(4) ChatGPT performs poorly in all studied types of vulnerabilities.

% \cn
{
Table~\ref{tab:rq2-2} shows the performance difference of each method on Top-25 most dangerous CWE.
By observing these results, we conclude that:
(1)  Overall, sequence-based methods perform better, especially SVulD, which shows their potentiality in practical usage.
(2) As for graph-based models, Devign (i.e., 397) outperforms Reveal (i.e., 30) and IVDdetect (i.e., 380) with an improvement of 95 and 17 functions correctly classified, respectively.
As for the Top 5 dangerous CWEs, Devign also performs better, which shows the priority among other graph-based models.
(3) As for sequence-based models, SVulD (i.e., 415) performs best and improves LineVul (i.e., 394) by 21.
The powerful ability of SVulD is also consistent in the results of Top-5 dangerous CWEs.
}

% \vspace{-0.2cm}
\intuition{
\textbf{Finding 2}: 
(1) Different models have their own advantages in detecting different types of vulnerabilities.
Particularly, sequence-based models are skilled in ``Input Validation'', but graph-based models have a wide range (``API Abuse'', ``Input Validation'' and ``Security Features'').
(2) Generally, sequence-based models especially SVulD perform better and SVulD shows its promising potentiality in practical usage when detecting the most dangerous vulnerabilities.
}
% \vspace{-0.2cm}

\noindent
{$\bullet$ \bf [RQ-3]: \ul{Are Large Language Models capable of detecting vulnerabilities?}}
\label{sec:rq3}

\noindent
\textbf{Objective}.
Large Language Models (LLMs)~\cite{brown2020language} have been widely adopted since the advances in Natural Language Processing (NLP) which enable LLM to be well-trained with both billions of parameters and billions of training samples, and consequently brings a large performance improvement on various tasks.
LLMs can be easily used for a downstream task by being prompted~\cite{liu2023pre} and they can capture different knowledge from various domain data.
Previous studies~\cite{liu2021makes, lu2021fantastically} have shown that the strength of LLMs may vary widely depending on the prompts. 
Therefore, we aim to investigate how LLMs perform in detecting vulnerabilities across different prompt settings since no study has been conducted comprehensively on this topic.

\noindent
\textbf{Experimental Setup}.
We conduct experiments with the state-of-the-art LLM, ChatGPT~\cite{openai2022chatgpt}.
Besides, considering the consumption of interaction with ChatGPT caused by the large-scale dataset (i.e., 38,749 functions), we statistically sample from the testing dataset as suggested by previous work~\cite{croft2023data}, which can also reflect the target dataset as precise as possible.
In particular, we sample the instances with 95\% confidence and 3\% interval\footnote{https://surveysystem.com/sscalc.htm}.
Eventually, we obtain 1,039 instances to conduct our study.

Meanwhile, considering that different prompts will affect the performance of ChatGPT in vulnerability detection, we adopt three prompt settings for our study: (1) \textbf{Zero-Shot:} which directly prompts ChatGPT to detect vulnerabilities without providing any demonstrations, (2) \textbf{In-Context-Learning (ICL):} enables ChatGPT to directly generate an answer for vulnerability detection task by feeding a few prompted demonstrations (i.e. a few shots) as part of the input, and (3) \textbf{Chain-of-Thought (CoT):} prompts ChatGPT to achieve an answer after a step-by-step process, which largely improves performance on reasoning.
Studies have shown that CoT reasoning can be performed with zero-shot prompting (Zero-Shot CoT)~\cite{kojima2022large} or few-shot demonstrations (Few-Shot CoT)~\cite{wei2022chain}.
We consider five distinct strategies for selecting demonstrations to explore the influence of different demonstrations on Few-shot ICL and Few-shot CoT.
The details of five selection strategies are elaborated as follows.

\begin{itemize}
\item \textbf{Fixed Selection}. 
We select pre-set fixed demonstrations in a sequential order from up to six CWEs (i.e., CWE-416, CWE-476, CWE-79, CWE-200, CWE-20 and CWE-787) until limitation are reached.
These CWEs are selected from the Top-25 Most Dangerous Software Weaknesses.
\item \textbf{Random Selection}. 
We randomly select a few demonstrations from training data (i.e., cf. Section~\ref{sec:rq1} for details).
\item \textbf{Random$_{repo}$ Selection}.
We randomly select demonstrations from training data and these demonstrations are from the same projects that the target function belongs to.
\item \textbf{Diversity-based Selection}. 
We adopt a pre-trained model (i.e., CodeBERT~\cite{feng2020codebert}) to embed all the functions from training data and then uses K-means algorithm~\cite{macqueen1967some} for clustering with six centers. 
The demonstrations that are closest to each cluster center are selected to ensure diversity.
\item \textbf{Semantic-based Selection}. 
We utilize CodeBERT to embed all the functions from the training data as well as the target function. 
Subsequently, we select the most semantically similar demonstrations to the target function based on cosine similarity.
\end{itemize}

\begin{figure*}[htbp]
  \centering
  \includegraphics[width=\linewidth]{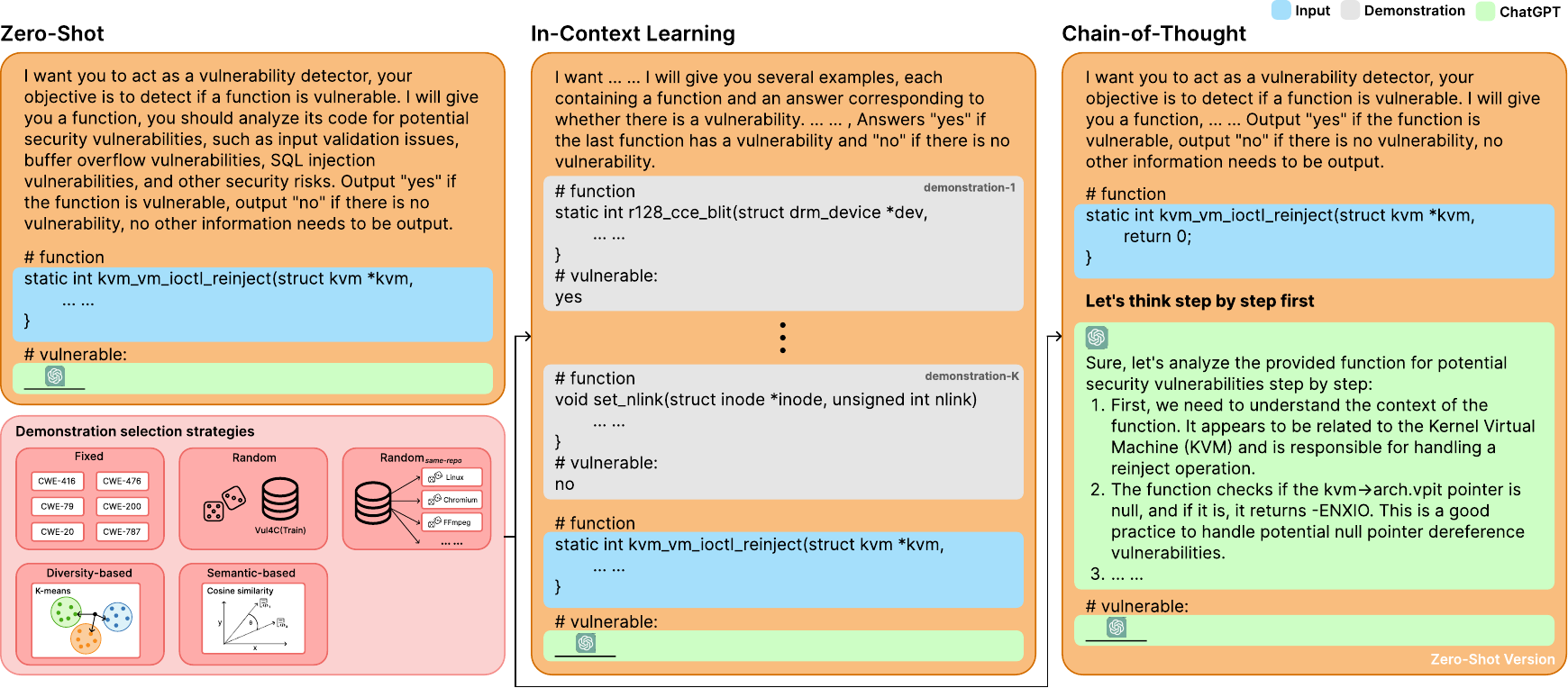}
  \caption{Vulnerability detection prompt templates used in our study}
  \label{fig:prompt_examples}
\end{figure*}

% \begin{figure*}[htbp]
%     \centering
%     \includesvg[width=.9\linewidth]{figs/llm_settings.svg}
%     \caption{Vulnerability detection prompt templates used in our study}
%     \label{fig:prompt_examples}
% \end{figure*}

% \sly
{
Fig.~\ref{fig:prompt_examples} presents examples of prompt templates under three different prompt settings. These prompt templates start with the instruction \texttt{``I want you to act as a vulnerability detector. Your objective is to detect... Output `yes' if the function is vulnerable..."} which explicitly states the task to be completed by the LLM and the expected output format.
If the prompt includes additional demonstrations (i.e., in few-shot setting), \texttt{``I will give you several examples..."} will also be inserted into the instruction to explicitly indicate to the LLM the presence of multiple functions and their vulnerability detection results within the prompt.
In the few-shot setting, we employ the effective and efficient selection strategy mentioned above to choose as many demonstrations as possible from the training set until we reach the LLM's maximum input window token limitation (i.e., 4,096). Including more demonstrations in the prompt can convey task-specific knowledge to LLMs through the correlation between input and output~\cite{min-etal-2022-rethinking}, thus enhancing LLM performance.
More specifically, in the ICL setting, we employ five selection strategies. For CoT, we manually craft the reasoning process for each demonstration, and hand-crafted reasoning is superior to LLM-generated reasoning~\cite{kojima2022large}. Therefore, we only consider the Few-Shot setting combined with two selection strategies (i.e., Fixed Selection and Diversity-based Selection).
In the Zero-Shot Cot setting, we use \texttt{``Let's think by think"} to prompt the LLM to generate its own reasoning process.
These demonstrations and reasonings are incorporated into the prompt in a specific format, as denoted by the gray background in the figure. Subsequently, the function to be detected for vulnerabilities is added to the end of the template, forming the resulting prompt that instructs LLM to produce the final detection result.
}

Overall, we explore nine prompt designs for ChatGPT when detecting vulnerability and the details of the setting are illustrated in Table~\ref{tab:rq3}.

% Table generated by Excel2LaTeX from sheet 'rq3'
\begin{table*}[htbp]
  % \vspace{-0.2cm}
  \centering
  \caption{The prompt design for ChatGPT when detecting vulnerability}
  \resizebox{.8\linewidth}{!}{
  \begin{threeparttable}
      
    \begin{tabular}{cccccccccccc}
    \toprule
    \multirow{2}[4]{*}{\textbf{\tabincell{c}{Zero-Shot}}} & \multirow{2}[4]{*}{\textbf{\tabincell{c}{In-Context Learning}}} & \multirow{2}[4]{*}{\textbf{\tabincell{c}{Chains-of-Thoughts}}} & \multicolumn{5}{c}{\textbf{Example Selection Strategy}} & \multirow{2}[4]{*}{\textbf{Accuracy}} & \multirow{2}[4]{*}{\textbf{Recall}} & \multirow{2}[4]{*}{\textbf{Precision}} & \multirow{2}[4]{*}{\textbf{F1}} \\
    \cmidrule{4-8}          &       &       & \textbf{Fixed} & \textbf{Random} & \textbf{Random$_{repo}$} & \textbf{Diversity} & \textbf{Semantic} &       &       &       & 
    
    \\

%  \multirow{2}[4]{*}{\textbf{\tabincell{c}{ZoSt}}} & \multirow{2}[4]{*}{\textbf{\tabincell{c}{ICL}}} & \multirow{2}[4]{*}{\textbf{\tabincell{c}{CoT}}} & \multicolumn{5}{c}{\textbf{Example Selection Strategy}} & \multirow{2}[4]{*}{\textbf{Accuracy}} & \multirow{2}[4]{*}{\textbf{Recall}} & \multirow{2}[4]{*}{\textbf{Precision}} & \multirow{2}[4]{*}{\textbf{F1}} \\
% \cmidrule{4-8}          &       &       & \textbf{Fixed} & \textbf{Rdm} & \textbf{Rdm$_{repo}$} & \textbf{Div} & \textbf{Sem} &       &       &       & 

% \\
    \midrule
    \ding{51}     &       &       &       &       &       &       &       & \cellcolor[rgb]{ .886,  .937,  .855}\textbf{0.977} & \cellcolor[rgb]{ .996,  .78,  .812}0 & \cellcolor[rgb]{ .996,  .78,  .812}0 & \cellcolor[rgb]{ .996,  .78,  .812}0 \\
    \midrule
          & \ding{51}     &       & \ding{51}     &       &       &       &       & 0.960 & 0.042 & 0.050 & 0.046 \\
          & \ding{51}     &       &       & \ding{51}     &       &       &       & 0.934 & 0.042 & 0.021 & 0.028 \\
          & \ding{51}     &       &       &       & \ding{51}     &       &       & 0.932 & 0.125 & \cellcolor[rgb]{ .886,  .937,  .855}\textbf{0.057} & \cellcolor[rgb]{ .886,  .937,  .855}\textbf{0.078} \\
          & \ding{51}     &       &       &       &       & \ding{51}     &       & 0.921 & 0.042 & 0.017 & 0.024 \\
          & \ding{51}     &       &       &       &       &       & \ding{51}     & 0.946 & 0.042 & 0.029 & 0.035 \\
    \midrule
    \ding{51}     &       & \ding{51}     &       &       &       &       &       & 0.867 & 0.083 & 0.017 & 0.028 \\
          &       & \ding{51}     & \ding{51}     &       &       &       &       & 0.961 & \cellcolor[rgb]{ .996,  .78,  .812}0 & \cellcolor[rgb]{ .996,  .78,  .812}0 & \cellcolor[rgb]{ .996,  .78,  .812}0 \\
          &       & \ding{51}     &       &       &       & \ding{51}     &       & 0.733 & \cellcolor[rgb]{ .886,  .937,  .855}\textbf{0.375} & 0.033 & 0.061 \\
    \bottomrule
    \end{tabular}%
% $^\ast$ ``ZoSt'': Zero-Shot; ``ICL'': In-Context Learning; ``CoT'': Chains-of-Thoughts;
% ``Rdm'': Random; ``Div.'': Diversity; ``Sem'': Semantic
     \end{threeparttable}

}
  \label{tab:rq3}%
    % \vspace{-0.2cm}
\end{table*}%

\noindent
\textbf{Results}.
Table~\ref{tab:rq3} shows the comparison results among different prompt designs for ChatGPT when detecting vulnerability.
From the detailed results, we can achieve the following observations: 
(1) Different prompt setting results in varying performances and no one setting can achieve the best performs for all metrics.
(2) Overall, the combination of CoT and Diversity-based Selection achieved the best performance in terms of \textit{Recall} (i.e., 0.375), improving other setting a lot (i.e.,  $\leq 0.125$).
(3) ChatGPT prompted with in-context learning as well as \textit{Random$_{repo}$} strategy performance well in term of \textit{Precision} (i.e., 0.057) and \textit{F1} (i.e., 0.078) and also achieve a performance of \textit{Recall} (i.e., 0.125).
It seems to indicate that demonstrations from some domain with target function may help ChatGPT better to address the similar task.
(4) Though ``Zero-Shot'' achieves best in terms of \textit{Accuracy}, it fully performs worst in terms of other three metrics.
We further analyze and find that in this setting, ChatGPT almost predicts all function as a clean one, It seems to has no ability to distinguish between clean and vulnerable ones, which is also confirmed by its performance on other three performance metrics.
(5) Considering the highly imbalanced dataset in practice (i.e., 3.6\% vulnerability in our dataset),  ChatGPT prompted with in-context learning as well as \textit{Random$_{repo}$} strategy is the best setting.

% % (1) Overall, Zero-shot prompt design achieves the highest accuracy (0.977).
% % ICL, Zero-Shot CoT, and CoT under Diversity-based Selection outperformed the Zero-Shot prompt design in the other three evaluation metrics.
% (2) ICL under Random$_{repo}$ Selection  has improved all other prompt designs by 0.007$\sim$0.057 on Precision and 0.017$\sim$0.078 on F1.
% (3) 

% \vspace{-0.2cm}
\intuition{
\textbf{Finding 3}: 
(1) ChatGPT has limited ability to be directly used to detect the vulnerability and different prompt designs will highly affect its performance.
(2) Overall, ChatGPT prompted with in-context learning as well as Random$_{repo}$  selection strategy performs the best in terms of \textit{Precision} and \textit{F1-score}  and achieves a relatively good performance in terms of \textit{Recall}.
}
% \vspace{-0.3cm}

\subsection{D2: Interpretation of Learning-based Models for Vulnerability Detection}

\noindent
{$\bullet$ \bf [RQ-4]: \ul{What source code information does the learning-based model focus on?  Do different types of learning models agree on similar important code features?}}

\noindent
\textbf{Objective}.
Vulnerability detection models should help developers understand how they make their predictions (i.e., identify vulnerable code patterns).
Therefore, it is meaningful to investigate whether different deep learning models make decisions based on specific types of statements and help the model better be understood.
For instance, the model might pay more attention to ``if'' statements when detecting input validation vulnerabilities. 

Additionally, different types of learning-based approaches (i.e., graph-based and sequence-based) may focus on varying types of information, and figuring out the difference of code features concerned by models can help to better improve their abilities.

\noindent
\textbf{Experimental Setup}.
To explain the types of statements the model focuses on, we need to obtain the score for each token in the source code. 
We employ different interpretability techniques to acquire precise scores of tokens based on the characteristics of the studied models.
For graph-based models (e.g., Devign~\cite{zhou2019devign} and IVDetect~\cite{li2021vulnerability}), we utilize GNNExplainer~\cite{ying2019gnnexplainer}, which provides scores for each edge in the constructed graph, and subsequently, we calculate the score for each node by aggregating the scores of all incoming edges. 
Besides, since a node may contain several tokens, we assign the score to each corresponding token.
As for the Reveal, it employs a two-stage architecture consisting of a GNN for learning feature vectors and a representation model for classification. 
We adopt DeepLift~\cite{shrikumar2019learning} for the representation model to unveil the contribution of each neuron to the final prediction.
For sequence-based models (e.g., LineVul~\cite{fu2022linevul} and SVulD~\cite{ni2023distinguishing}), we use the attention layer to get each tokens' score since they are Transformer-based model ~\cite{vaswani2023attention}, naturally providing reasoning behind the prediction decision~\cite{serrano2019attention}.

% \sly{
% In order to explain the types of statement the model focuses on, we first need to obtain the score for each token in the source code. Depending on the type and structure of the model, we have employed different interpretability techniques to acquire more precise scores.
% For graph models such as Devign\cite{zhou2019devign} and IVdetect\cite{li2021vulnerability}, we utilized GNNExplainer\cite{ying2019gnnexplainer}. GNNExplainer provides scores for each edge in the graph, and subsequently, we calculate the score for each node by aggregating the scores of all incoming edges. Additionally, since a node may correspond to multiple tokens, we assign score to each corresponding token.
% As for the Reveal, it employs a two-stage architecture consisting of GNN and representation learning model. The feature vectors learned by the GNN are used for subsequent vulnerability classification by the representation model. Therefore, we employed DeepLift\cite{shrikumar2019learning} for the representation model to unveil the contribution of each neuron to the final prediction.
% Regarding LineVul\cite{fu2022linevul} and SVulD\cite{ni2023distinguishing}, both rely on Transformer\cite{vaswani2023attention} as backbone, and its attention mechanism naturally provides insights into the reasoning behind the predictions of that model\cite{serrano2019attention}, so we use the attention layer to get score for each token.
% }

\begin{table}[htbp]
% \vspace{-0.3cm}
  \centering
  \caption{Types of Statements}
  \resizebox{\linewidth}{!}
  {
    \begin{tabular}{l|l}
    \toprule
\textbf{Statement Type} & \textbf{Brief Description}\\
\midrule
If Statement & \textit{if} keyword and condition expression\\
For Statement & \textit{for} keyword, initialization, condition, iteration expression\\
While Statement & \textit{while} keyword and condition expression\\
Jump Statement & \textit{goto}, \textit{break}, \textit{continue}\\
Switch Statement & \textit{switch} keyword and condition expression\\
Case Statement & \textit{case},\textit{default} keyword and value expression  \\
Return Statement & \textit{return} keyword\\
Arithmetic Operation & +, -, *, /, \% Binary expression  \\
Relational Operation & ==, !=, >, <, >=, <= Binary expression \\
Logical Operation & $\&\&$ , $||$ Binary expression\\
Bitwise Operation & $\&$, $|$, $\land$ , $<<$, $>>$ Binary expression \\
Declaration Statement & variable type and name\\ 
\bottomrule
\end{tabular}
}
% \vspace{-0.3cm}
\label{tab:rq3-2}%
\end{table}

After obtaining scores for each token, we can obtain the score for each line by summing up the scores of all tokens within it.
For each correctly classified vulnerable function in the testing dataset, we select the top 10 lines with the highest scores and treat them as the most important code features contributing to the model's decision.
Subsequently, we utilize Tree-sitter~\cite{tree-sitter} to parse 15 types of statements (shown in Table~\ref{tab:rq3-2}) within the functions and count the occurrences of each statement type among the top 10 lines.

% \sly{
% After obtaining scores for each token, we calculate the score for each code line by summing up the scores of all tokens within that line. 
% For each vulnerable example in the test set that are classified correctly, we select the top 10 lines with the highest scores, considering them as the most important features contributing to the model's classification results.
% Subsequently, we utilize Tree-sitter\cite{tree-sitter} to parse 15 types of statements(Details can be found in Table \ref{tab:rq3-2}) within the functions and count the occurrences of each statement type among the top 10 lines.
% }

We also apply \textit{t}-SNE~\cite{JMLR:v9:vandermaaten08a}, a visualization technology mapping high-dimensional features into two-dimensional features, to explore the separability of studied models between vulnerable functions and non-vulnerable functions.
For a better illustration, we randomly select 10,000 examples from the testing dataset and extract the hidden vectors before making the final binary classification decision as the high-dimensional features for different models, e.g., the hidden vector of the \texttt{[CLS]} used for sequenced-based models and the hidden features of each node used for graph-based models.

% \sly{
% We apply t-SNE\cite{JMLR:v9:vandermaaten08a} technique to explore the separability of the feature vectors obtained during the training process of the current vulnerability model. t-SNE can map high-dimensional feature data to a two-dimensional space, enabling us to conduct visual analysis and intuitively assess the performance of the existing model.
% We randomly selected 10,000 examples from the test set for the t-SNE visualization. 
% For different models, we extract the hidden vectors before making the final binary classification decision as the dimensional reduction features for t-SNE. 
% For instance, in the case of sequence models, we choose to use the hidden vector of the [CLS] token as input, while for graph models, we use the hidden features of each node as input.
% }

% \begin{figure}[htbp]
% \vspace{-0.3cm}
%     \centering
%     \includesvg[width=.75\linewidth]{figs/statement_types.svg}
%     \caption{Number of occurrences for each statement type in the Top 10 most probable vulnerability lines. (diagonal shadow indicates sequence-base models)}
%     \label{fig:statement_types}
% \end{figure}

\begin{figure}[htbp]
  % \vspace{-0.3cm}
      \centering
      \includegraphics[width=\linewidth]{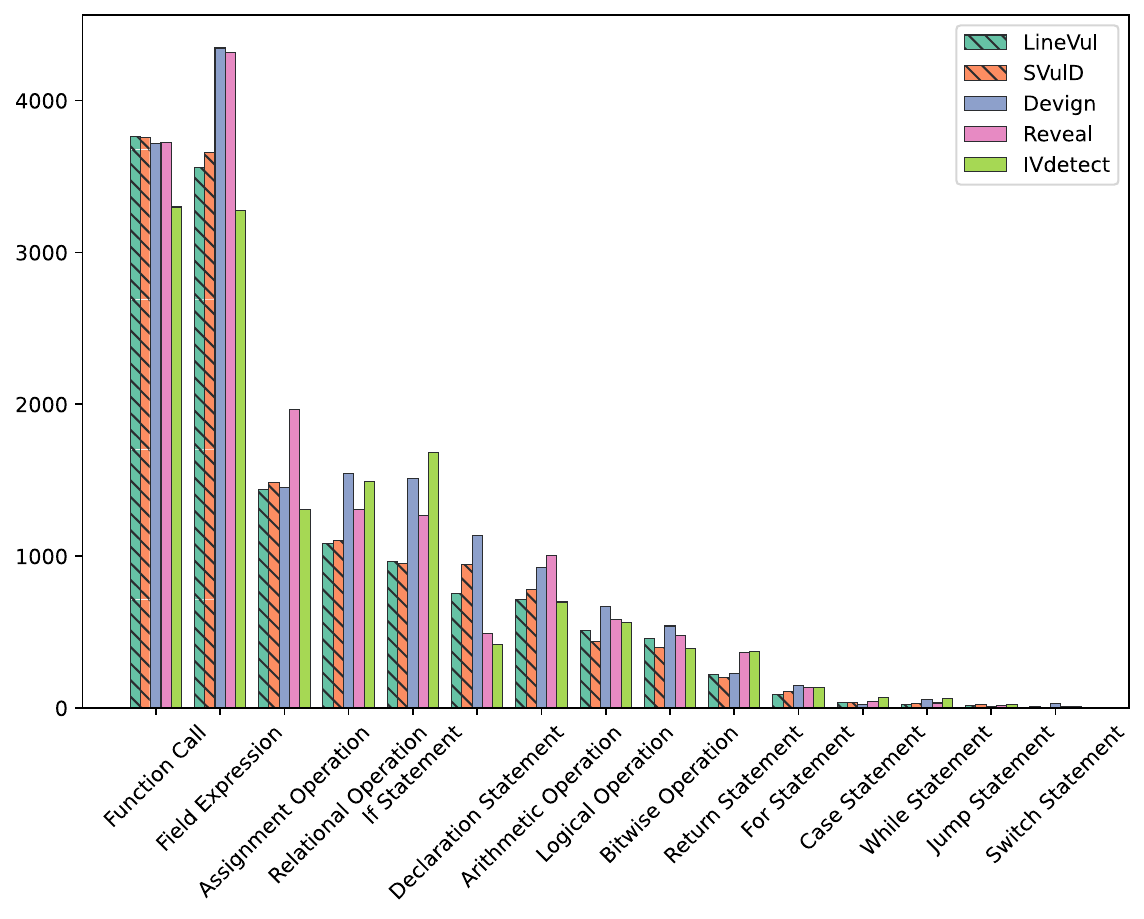}
      \caption{Number of occurrences for each statement type in the Top 10 most probable vulnerability lines. (diagonal shadow indicates sequence-base models)}
      \label{fig:statement_types}
  \end{figure}

% \begin{table}[htbp]
% \vspace{-0.3cm}
%   \centering
%   \caption{Types of Statements}
%   \resizebox{\linewidth}{!}{
%     \begin{tabular}{l|p{6cm}}
%     \toprule
% \textbf{Statement Type} & \textbf{Brief Descrption}\\
% \midrule
% If Statement & \textit{if} keyword and condition expression\\
% For Statement & \textit{for} keyword, initialization, condition, iteration expression\\
% While Statement & \textit{while} keyword and condition expression\\
% Jump Statement & \textit{goto}, \textit{break}, \textit{continue}\\
% Switch Statement & \textit{switch} keyword and condition expression\\
% Case Statement & \textit{case},\textit{default} keyword and value expression  \\
% Return Statement & \textit{return} keyword\\
% Arithmetic Operation & +, -, *, /, \% Binary expression  \\
% Relational Operation & ==, !=, >, <, >=, <= Binary expression \\
% Logical Operation & $\&\&$ , $||$ Binary expression\\
% Bitwise Operation & $\&$, $|$, $\land$ , $<<$, $>>$ Binary expression \\
% Declaration Statement & variable type and name\\ 
% \bottomrule
% \end{tabular}
% }\vspace{-0.3cm}
% \label{tab:rq3-2}%
% \end{table}

\begin{figure}[htbp]
  % \vspace{-0.3cm}
      \centering
      \includegraphics[width=\linewidth]{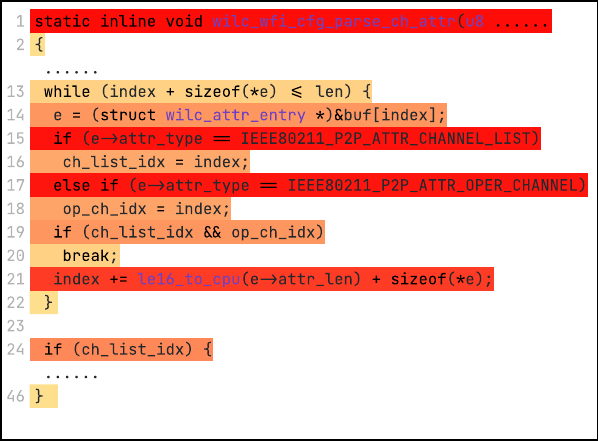}
      \caption{LineVul interpretation result(CVE-2022-47519\cite{CVE-2022-47519}) }
      % \vspace{-0.3cm}
      \label{fig:cve_example}
  \end{figure}

\noindent
\textbf{Results}.
Fig.~\ref{fig:statement_types} shows the results and we obtain the following findings:
(1) ``Function Call'' and ``Field Expression'' are the most risky operations identified by both graph-based and sequenced-based models.
The operating frequency of the two statement types exceeds 50\% among the studied 15 different statement types, which seems that both operations will introduce unstable factors to functionality and are prone to introduce vulnerabilities. For example, 
{Fig.~\ref{fig:cve_example} shows a function from the Linux project that aims to parse the channel attribute in the WIFI configuration.}
The code (Line 21)  makes a function call (i.e., ``\textit{le16\_to\_cpu}'') and brings a risk to the current function.
That is, the external function should not be called directly for another operation, and further input validation is required to ensure the attribute has enough space to avoid ``\textit{out-of-bounds write}'' vulnerability.
(2) 
% Condition statements是条件语句 不对
% Condition statements (i.e., ``for'', ``case'', ``while'', ``jump'', and ``switch'') are relatively paid low attention.
The models exhibit relatively low attention towards ``for'', ``case'', ``while'', ``jump'', and ``switch'' statement types.
% We conduct a manual analysis of these functions and find that defects/vulnerabilities existing in conditional statements can be easily detected during debugging and compilation. 
We conducted a manual analysis of these functions and found that the statements are generally simple, making them less prone to vulnerabilities. 
For example, the ``\textit{while}'' condition statement (Line 13 in Fig.~\ref{fig:cve_example}) is 
straightforwardly presented, and developers can easily identify the termination criteria while writing the program.
% Furthermore, these statement types themselves are not particularly prone to vulnerabilities; for example, "case" and "switch" statements usually involve a straightforward selection of branches based on a single variable's value.
(3) {Sequence-based models (i.e., LineVul and SVulD) perform similarly on different types of statements, possibly because both of them are built upon CodeBERT}~\cite{fengetal2020codebert} and variants~\cite{guo2022unixcoder}.
For example, for each type of statement, the two methods seem to achieve the same attention numbers.
% \chao{need more detail about the trends}
(4) Graph-based models pay varying attention to different types of statements.
For example, for ``Field Expression'', both Reveal and Devign pay more attention than IVDetect.
For ``Assignment Operation'', the Reveal pays more attention than both Devign and IVDetect.
The difference may be caused by the way to encode the internal node among graph-based models.
IVDetect strives to encode as much information as possible from a single line of code (e.g., AST, CDG, etc.) into a single node, Devign directly utilizes nodes generated by \textit{Joern} as the nodes presented in the graph, which may explain why IVDetect shows less sensitive to the operation of accessing or operating members in \textit{class} or \textit{struct}, since IVDetect merges multiple field expressions into a single node, losing the detailed information, and consequently reduces its attention to such statement type.

% \begin{figure}[htbp]
% \vspace{-0.3cm}
%     \centering
%     \includesvg[width=.75\linewidth]{figs/CVE-2022-47519.svg}
%     \caption{LineVul interpretation result(CVE-2022-47519\cite{CVE-2022-47519}) }
%     \vspace{-0.3cm}
%     \label{fig:cve_example}
% \end{figure}

\begin{figure*}[htbp]
% \vspace{-0.3cm}
\centering
\subfigure[Devign]{
\begin{minipage}[t]{0.19\linewidth}
\centering
\includegraphics[width=1.35in]{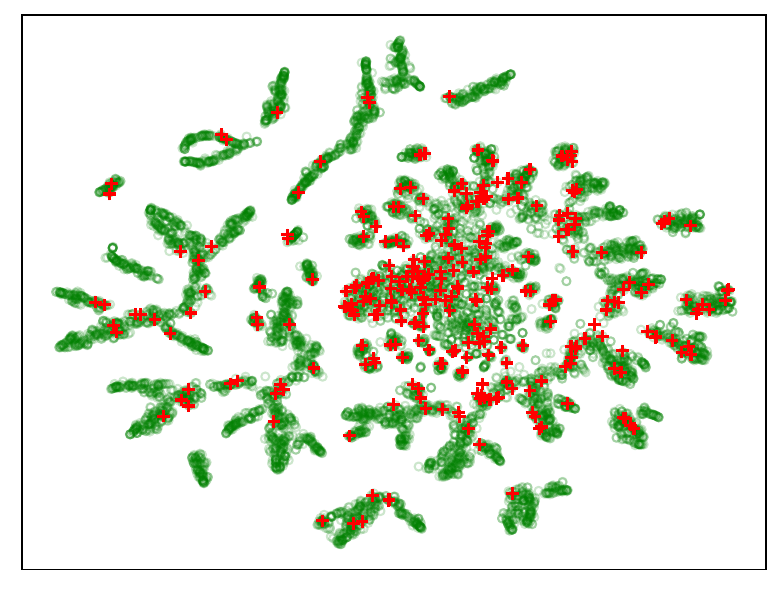}
%\caption{fig1}
\end{minipage}%
}%
\subfigure[{\sc ReVeal}]{
\begin{minipage}[t]{0.19\linewidth}
\centering
\includegraphics[width=1.35in]{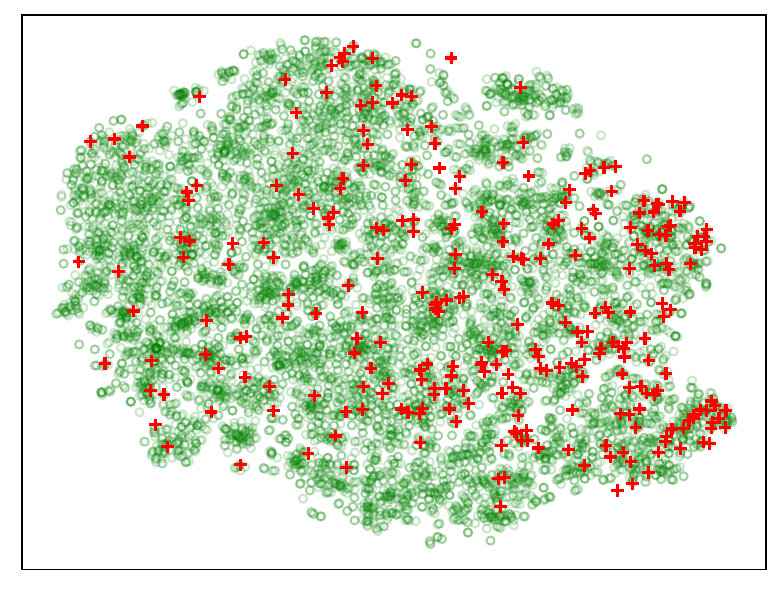}
%\caption{fig2}
\end{minipage}%
}%
\subfigure[{\sc IVDetect}]{
\begin{minipage}[t]{0.19\linewidth}
\centering
\includegraphics[width=1.35in]{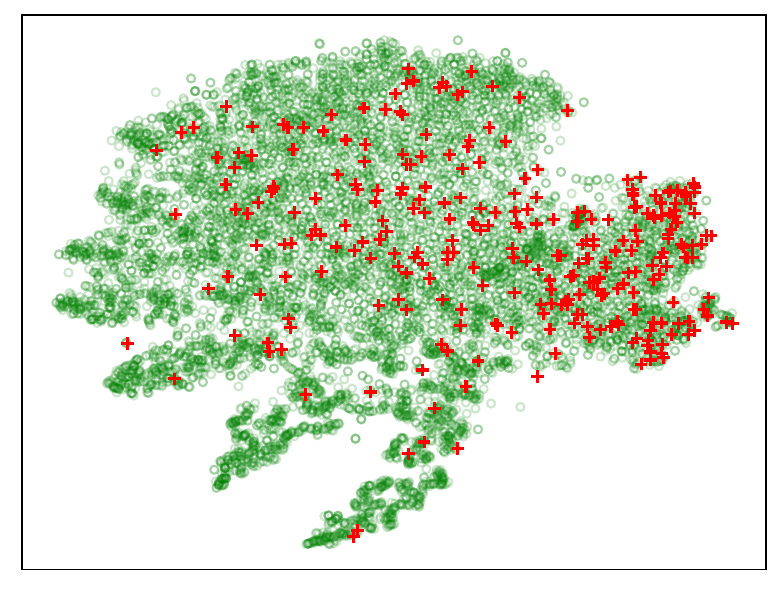}
%\caption{fig2}
\end{minipage}
}%
\subfigure[LineVul]{
\begin{minipage}[t]{0.19\linewidth}
\centering
\includegraphics[width=1.35in]{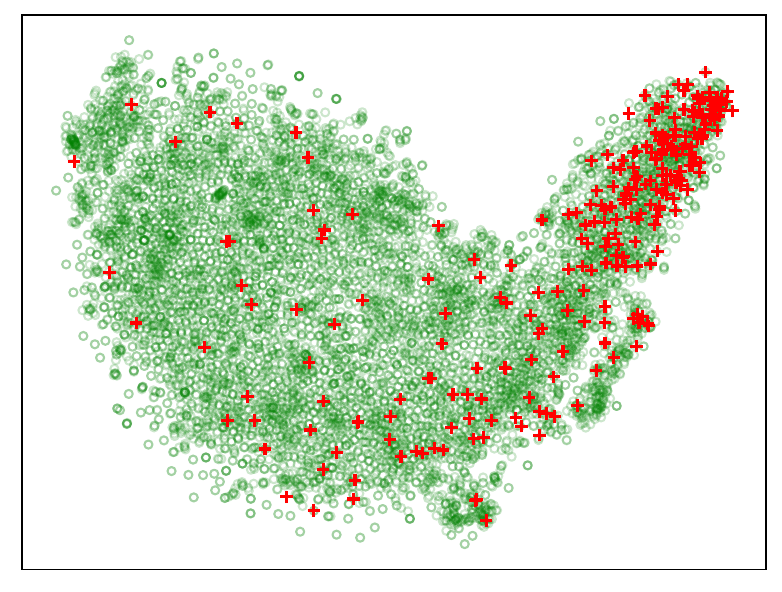}
%\caption{fig2}
\end{minipage}
}%
\subfigure[SVulD]{
\begin{minipage}[t]{0.19\linewidth}
\centering
\includegraphics[width=1.35in]{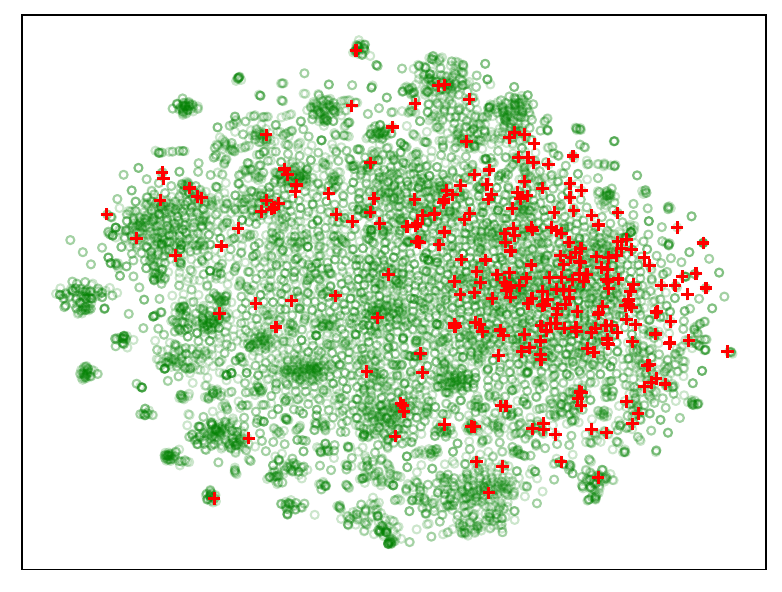}
%\caption{fig2}
\end{minipage}
}%
\centering
\caption{Visualization of the separation between vulnerable (denoted by \textcolor{red}{+}) and non-vulnerable (denoted by \textcolor{green}{$\bigcirc$}).}
\label{fig:rq3-2-tsne}
% \vspace{-0.35cm}
\end{figure*}

Fig.~\ref{fig:rq3-2-tsne} illustrates the visualization of separating vulnerable functions from non-vulnerable functions and  we obtain the following observations:
(1) All the figures show an overlap between the functions with or without vulnerabilities, which means that all the learning-based models have limited ability to distinguish them.
By analyzing the types of statements that the models focus on (i.e., ``Function Call''),  it seems that the models need the context of the externally called functions to enrich the input information, which helps to better understand the functionality.
(2) Sequence-based models (i.e., LineVul and SVulD) seem to have a better separation boundary (i.e., more concentrated) than graph-based models, especially LineVul seems to perform best, which is also consistent with the results obtained in RQ-1.

% \sly{
% Figure~\ref{fig:rq3-2-tsne} illustrates the t-SNE visualization of feature vectors of the function code after processing by different models.
% From the visualization results, the following conclusions can be intuitively drawn.
% (1) In all of these visualizations, there is an overlap between the examples with and without vulnerabilities, and the models have not been able to effectively capture the key features between the vulnerable and non-vulnerable samples, leading to confusion between the classifications.
% From the analysis of the types of statements that the models focus on, we speculate that the models need the context of the externally called functions so that the input data contains rich discriminative information in order for the models to achieve more accurate vulnerability detection results.
% (2) Based on the observation of the t-SNE plots, we can clearly see that LineVul exhibits the best class separation ability among the other four models. It tightly clusters examples with vulnerabilities, forming distinct clustering structures. On the other hand, the other four models fail to demonstrate such evident class separation, as their vulnerable and non-vulnerable examples appear to be randomly scattered.
% This finding also corroborates the results in RQ1, where LineVul demonstrates superior performance in vulnerability detection.
% }

% \vspace{-0.2cm}
\intuition{
\textbf{Finding 4}: 
(1) Both graph-based and sequence-based methods technically focus on two types of statements: Function Calls and Field Expressions, which may involve vulnerable or incredible operations to functionality.
(2) The existing learning-based models still have limited ability to distinguish vulnerable functions from non-vulnerable functions.
Sequence-based models perform better than the graph-based models.
(3) Feeding external called function information sequence-based method could further improve sequence-based models'  ability.
}

\subsection{D3: Stability of Learning-based Models for Vulnerability Detection}

\noindent
{$\bullet$ \bf [RQ-5]: \ul{Do learning-based models agree on the vulnerability detection results with themselves when the input is insignificantly changed?}}

\noindent
\textbf{Objective}.
An optimal vulnerability detection model should base its decisions on the root cause of vulnerabilities while demonstrating robustness against the potential impact of code layout or unrelated noise.
Therefore, we want to assess the stability of the studied models by evaluating their generalizability to slightly modified but semantically equivalent input.

\noindent
\textbf{Experimental Setup}.
We apply four types of semantic-preserving transformations (introduction along with examples are shown in Table~\ref{tab:rq5-1}) 
to each testing sample in the original \datasetname dataset to construct four distinct variants of the test set.
% to each function in the original \datasetname dataset to construct four distinct variants.
\textbf{(1) Remove all comments}. For each function, we remove all the comments in its source code. 
% 2. Insert comment：随机在函数的函数体中插入单行注释，插入的注释行数=15%*方法体行数（注意：注释的内容不一定和该函数相关，插入注释的位置随机）
\textbf{(2) Insert comments}. For each function, we randomly insert single-line comments into its function body. 
The number of inserted comments equals 15\% of the total number of lines in the function body, and the insertion positions are randomly selected. 
Note that the content of the comments may not be relevant to the function.
% 3. Insert irrelevant code： 插入 不执行、不影响函数结果的代码，每个函数体只随机插入一条无关代码。
\textbf{(3) Insert irrelevant code}. 
We randomly insert a single line of unrelated code for each function, which will not influence the function's functionality and output.
% 4. Rename all identifier：将函数的参数、函数体内声明的变量名，全部替换为VAR0、VAR1...VARX的格式。
\textbf{(4) Rename all identifiers}. 
For each function, we consistently replace the names of its parameters and variables declared within its function body with \texttt{VAR0}, \texttt{VAR1}, $\cdots$, \texttt{VARX}, which ensures that the program semantics and functionalities remain unchanged.
% We adopt the same experimental setting in RQ-1.
% \hl{Then, we train and evaluate the models using the four variation datasets on the same train/valid/test partitions}.
Then, we test the models trained in Section~\ref{sec:rq1} on the four variant test sets.
We adopt a comprehensive performance metric F1 to analyze the performance difference.

\begin{table}[htbp]
  % \vspace{-0.2cm}
  \centering
  \caption{Semantic-preserving Transformation Types}
  \resizebox{\linewidth}{!}{
    \begin{tabular}{l|p{5cm}|l}
    \toprule
    \textbf{Transformation Type} & \textbf{Summary} & \textbf{Example} \\
    \midrule
    Remove all comments & Remove all comments from the function &\sout{/* Initializing variables before the main loop. */} \\
    Insert comments & Randomly insert comments into the function & /* A loop to iterate over elements in an array. */ \\
    Insert irrelevant code & Randomly insert unrelated code into the function & if(0) \{\} \\
    Rename all identifiers & Replace parameters and declared variables with VARX & int delta; \ding{232} int VAR2; \\
    \bottomrule
    \end{tabular}%
  }
\label{tab:rq5-1}%
  % \vspace{-0.2cm}
\end{table}

\begin{table}[htbp]
% \vspace{-0.3cm}
  \centering
  \caption{Performance difference between original \datasetname test set and its four semantically-equivalent variants}
  \resizebox{\linewidth}{!}{
    \begin{threeparttable}
        % Table generated by Excel2LaTeX from sheet '8.1update'
        \begin{tabular}{llrrrrr}
        \toprule
        \textbf{Types} & \textbf{Models} & \textbf{Original} & \textbf{\makecell{Remove All\\Comments}} & \textbf{\makecell{Insert\\Comments}} &\textbf{ \makecell{Insert Irrelevant\\ Code}} & \textbf{\makecell{Rename All\\ Identifiers}} \\
        \midrule
        \multicolumn{1}{l}{\multirow{3}[2]{*}{\textbf{\tabincell{l}{Graph\\Based}}}} & Devign & 0.122  & 0.075 (38.8\%$\downarrow$) & 0.075 (38.3\%$\downarrow$) & 0.073 (40.0\%$\downarrow$) & 0.075 (38.4\%$\downarrow$) \\
              & Reveal & 0.125  & 0.099 (20.7\%$\downarrow$) & 0.093 (25.6\%$\downarrow$) & 0.094 (24.5\%$\downarrow$) & 0.097 (22.0\%$\downarrow$) \\
              & IVdetect & 0.141  & 0.051 (64.0\%$\downarrow$) & 0.052 (63.0\%$\downarrow$) & 0.055 (61.2\%$\downarrow$) & 0.025 (82.5\%$\downarrow$) \\
        \midrule
        \multicolumn{1}{l}{\multirow{3}[4]{*}{\textbf{\tabincell{l}{Sequence\\Based}}}} & LineVul & 0.195  & 0.196 (0.6\%$\uparrow$) & 0.186 (4.5\%$\downarrow$) & 0.195 (0.1\%$\uparrow$) & 0.183 (6.3\%$\downarrow$) \\
              & SVulD & 0.172  & 0.174 (1.2\%$\uparrow$) & 0.163 (5.4\%$\downarrow$) & 0.179 (3.8\%$\uparrow$) & 0.186 (7.9\%$\uparrow$) \\
        \cmidrule{2-7}      & ChatGPT$^\ast$ & 0.078  & 0.073 (6.0\%$\downarrow$) & 0.068 (12.7\%$\downarrow$) & 0.089 (13.6\%$\uparrow$) & 0.066 (14.9\%$\downarrow$) \\
        \bottomrule
        \end{tabular}%
        $^\ast$ Notice that the performance of ChatGPT is calculated on statistical sampling with 95\% confidence.
    \end{threeparttable}
  }
\label{tab:rq5-2}%
\end{table}

% \vspace{-0.2cm}
\noindent
\textbf{Results}.
% ******* 在这个rq里liyu使用的ChatGPT的策略是(ICL-same-repo)，要不要在结果表格最下面*的注解里说明一下 ******
Table~\ref{tab:rq5-2} shows the results of the studied models on the original test set and its four variants.
From the results, we achieve the following observations.
(1) All studied models are unstable to the four types of semantic-preserving transformations.
(2) Sequence-based models achieve a relatively smaller performance change, which means that these models are more stable than graph-based models.
% This phenomenon 
% TODO: sequence模型稳定的原因（可能要再改改）
The robustness of the sequence-based models may be explained by their elaborate and complex model architectures with large amounts of parameters and also indicates the existence of less meaningful tokens in code~\cite{zhang2022diet}.
(3) 
% Graph-based models are heavily affected by input modifications with a performance decrease of 20.7\%$\sim$82.5\%.
Graph-based models suffer from a severe performance decrease.
% TODO: 解释一下图模型为啥不稳定？
In particular, IVDetect is affected the most, whose performance drops by 61.2\%$\sim$82.5\%.
(4) Overall, ``Rename all identifiers'' impacts more to models' stability than other types of transformations.

\intuition{
\textbf{Finding 5}:
% All learning-based models are not stable to the change to input even the changes are semantically-equivalent.
% Overall, graph-based models are a little more stable to subtle input changes than sequence-based models.
All the learning-based models are unstable to input changes even if these changes are semantically equivalent.
Sequence-based models are more stable to subtle input changes than graph-based models.
}
% \vspace{-0.3cm}

\subsection{D4: Ease of Use of Learning-based Models for Vulnerability Detection}

\noindent
{$\bullet$ \bf [RQ-6]: \ul{What types of efforts should be paid before using a model?
In what scenarios can learning-based models be applied?}}

\noindent
\textbf{Objective}.
% \xxd
{We want to assess the ease of use of the vulnerability detection models by examining their input requirements and model features. These aspects can offer valuable insights for practitioners who seek practical applications of these models.
}

\noindent
\textbf{Experimental Setup}.
% \xxd
{
We carefully document the key steps for reproducing the graph-based, sequence-based models and ChatGPT.
Specifically, we verify the input requirements for each model by examining its requirement of program integrity (i.e., whether it can handle incomplete input programs), 
compilation (i.e., whether the input program needs to be compiled), 
and input size (i.e., the upper limit of input).
Furthermore, during training and inference, we record for each model whether it requires fine-tuning to ensure its optimal performance,
whether its source code is available,
the minimum hardware requirement,
the configuration difficulty,
and the data privacy security level.
}
% \sly{
% In order to investigate the usability of vulnerability detection models, we carefully documented the key steps required to reproduce relevant graph models, sequence models, and ChatGPT. Specifically, we verified whether the input programs needed compilation and assessed the model's ability to handle incomplete input programs.
% Furthermore, during training and inference, we record the minimum hardware specifications necessary for the model and evaluate the ease of secondary configuration of the code and the simplicity of the configuration process to help other researchers to be able to easily use and customize the model.
% }

\begin{table*}[htbp]
% \vspace{-0.3cm}
  \centering
  \caption{Ease of Use of Learning-based Models}
  
\resizebox{\linewidth}{!}{
\begin{tabular}{lcccccccc}
\toprule
\multicolumn{1}{l}{\bf Models} & \multicolumn{3}{c}{\textbf{Input Requirements}}                                                                   & \multicolumn{5}{c}{\textbf{Model Features} }                                                                                                                                                              \\
\cmidrule(lr){2-4}
\cmidrule(lr){5-9}
\multicolumn{1}{l}{} & \multicolumn{1}{c}{Program Integrity} & \multicolumn{1}{c}{Compilation} & \multicolumn{1}{c}{Input Size} & \multicolumn{1}{c}{Fine-Tuning} & \multicolumn{1}{c}{Code Availability} & \multicolumn{1}{c}{Hardware Requirement} & \multicolumn{1}{c}{Configuration Difficulty} & \multicolumn{1}{c}{Privacy} \\
    \midrule
Devign & \ding{51}& \ding{55} & Medium & \ding{51} & \ding{51}& >1GB & Difficult & Safe \\
Reveal &  \ding{51}& \ding{55} & Medium &\ding{51} &\ding{51} &  >1GB &Difficult & Safe\\
IVDetect & \ding{51} &\ding{55} & Medium&\ding{51} &\ding{51} & >1GB &Difficult & Safe\\
LineVul & \ding{55} &\ding{55} & Small &\ding{51} & \ding{51}&  >6GB&Medium & Safe\\
SVulD & \ding{55}& \ding{55}& Small& \ding{51} &\ding{51} &  >6GB & Medium& Safe\\
ChatGPT &\ding{55} &\ding{55} & Large & \ding{55} & \ding{55}&  API &Easy & Unsafe\\
\bottomrule
\end{tabular}}
% \vspace{-0.3cm}
\label{tab:rq6}
\end{table*}

\noindent
\textbf{Results}.
% \cn{analyze obtained results}
% \xxd
{
We summarize the ease of use of the models in Table~\ref{tab:rq6}.
From the results, we obtain the following conclusions:
(1) Graph-based models require complete input programs since their inputs must be successfully parsed into graphs, while sequence-based models do not require program integrity.
(2) None of the models require the input programs to be compilable.
(3) ChatGPT has the largest input size ($\leq$16K tokens), while sequence-based models LineVul and SVulD are limited to a small input size ($\leq$ 512 tokens). The input size of graph-based models is medium.
(4) All the models, except for ChatGPT, have released their implementation code and require fine-tuning to enhance the performance, while ChatGPT is closed-source and hard to fine-tune.
(5) ChatGPT is the most user-friendly method, as it can be used directly through API or on a website. 
Graph-based models demand small memory (>1GB) but require complex preprocessing steps and configurations to construct code graphs, while sequence-based models require larger memory (>6GB) but involve only a small amount of coding work.
(6) All models, except for ChatGPT, are privacy-safe as they can be deployed on the user's own server, while ChatGPT carries a potential risk of privacy leakage.
}

% \begin{table*}[htbp]\vspace{-0.3cm}
%   \centering
%   \caption{Ease of Use of Learning-based Models}
  
% \resizebox{\linewidth}{!}{
% \begin{tabular}{lcccccccc}
% \toprule
% \multicolumn{1}{l}{\bf Models} & \multicolumn{3}{c}{\textbf{Input Requirements}}                                                                   & \multicolumn{5}{c}{\textbf{Model Features} }                                                                                                                                                              \\
% \cmidrule(lr){2-4}
% \cmidrule(lr){5-9}
% \multicolumn{1}{l}{} & \multicolumn{1}{c}{Program Integrity} & \multicolumn{1}{c}{Compilation} & \multicolumn{1}{c}{Input Size} & \multicolumn{1}{c}{Fine-Tuning} & \multicolumn{1}{c}{Code Availability} & \multicolumn{1}{c}{Hardware Requirement} & \multicolumn{1}{c}{Configuration Difficulty} & \multicolumn{1}{c}{Privacy} \\
%     \midrule
% Devign & \ding{51}& \ding{55} & Medium & \ding{51} & \ding{51}& >1GB & Difficult & Safe \\
% Reveal &  \ding{51}& \ding{55} & Medium &\ding{51} &\ding{51} &  >1GB &Difficult & Safe\\
% IVDetect & \ding{51} &\ding{55} & Medium&\ding{51} &\ding{51} & >1GB &Difficult & Safe\\
% LineVul & \ding{55} &\ding{55} & Small &\ding{51} & \ding{51}&  >6GB&Medium & Safe\\
% SVulD & \ding{55}& \ding{55}& Small& \ding{51} &\ding{51} &  >6GB & Medium& Safe\\
% ChatGPT &\ding{55} &\ding{55} & Large & \ding{55} & \ding{55}&  API &Easy & Unsafe\\
% \bottomrule
% \end{tabular}}\vspace{-0.3cm}
% \label{tab:rq6}
% \end{table*}

% \chao{consider privacy}
% \vspace{-0.2cm}
\intuition{
\textbf{Finding 6}:
Graph-based models require complete input programs and complex configurations to construct code graphs, while sequence-based models are easier to deploy.
Except for ChatGPT, all current models are relatively limited by input sizes, require fine-tuning to achieve enhanced performance, and are open-source and privacy-safe.
ChatGPT is the most user-friendly option regarding input requirements and model configurations, but it presents a potential risk of privacy leakage.
}
% \vspace{-0.3cm}

\subsection{D5: Economy Impact of Learning-based Models for Vulnerability Detection}

\noindent
{$\bullet$ \bf [RQ-7]: \ul{What are the costs caused by models from both time and economic aspects?}}

\noindent
\textbf{Objective}.
% \xxd
{
Deploying vulnerability models in a real-world setting requires appropriate resource allocation to ensure high cost-effectiveness.
Users are often interested in factors such as the effort required for model training and deployment, the model's processing speed for incoming requests, and the budget associated with the deployment.
Therefore, in this RQ, we aim to assess the time and economic costs of the models.
}

\noindent
\textbf{Experimental Setup}.
% \sly
{
% To fairly measure the time cost differences among all models and tools, 
We conduct experiments on a server with a uniform configuration equipped with an Intel(R) Xeon(R) Platinum 8358P CPU @ 2.60GHz, 755GB of RAM, and 10 NVIDIA GeForce RTX 3090 graphics cards.
During the data preprocessing phase, we utilize the tools Glove, Word2Vec, and Joern.
Glove and Word2Vec need to train on the train set, while Joern needs to extract graph information for all functions in the dataset. 
We adopt the latest versions of these tools available on GitHub.
% To avoid additional time costs resulting from inter-card communication and to provide more valuable references for future researchers, 
We decided to perform model training and inference on a single RTX 3090 graphics card and adjust the batch size to maximize the use of the GPU memory.
% The training and inference of deep learning models may be affected by other online users of the server utilizing additional hardware resources and the potential impact of ChatGPT API calls depending on the running status and user volume on the ChatGPT servers.
% To mitigate the chance factors mentioned above, 
We execute the experiments three times and calculate the average running time results to mitigate the bias.
We use the API provided by PyTorch to iteratively obtain the parameter size of the models.
The inference cost of using ChatGPT is calculated according to the pricing strategy provided on the OpenAI~\cite{openai-pricing} official website, and the version of ChatGPT we used is GPT-3.5 Turbo with a 4K context window.
For the other deep learning models, we calculate the cost of going from preprocess data to inference whole test set on hourly pricing using AWS's g5.xlarge instance~\cite{aws_g5}, which utilizes an NVIDIA A10G with similar performance to the NVIDIA 3090.
}

% \usepackage[normalem]{ulem}
% \begin{table}[htbp]
%   \centering
%     \caption{Time and economic costs of the models}
%     \resizebox{\linewidth}{!}{
% \begin{tabular}{lrrrrc}
%          \toprule
%          \textbf{Model} & \textbf{Preprocess Data Time} & \textbf{Training Time} & \textbf{Inference Time} & \textbf{Model Parameter} & \textbf{Cost} \\
% \midrule
% Devign   & 7,103s                & 2,836s         & 101s           & 0.97M           & 2.8056\$       \\
% Reveal   & 7,103s                & 5,220s         & 148s           & 1.09M           & 4.4378\$       \\
% IVDetect & 3,563s                & 1,3602s        & 916s           & 1.01M           & 4.8821\$       \\
% LineVul  & 0s                   & 13,274s        & 322s           & 124.65M         & 6.1712\$       \\
% SVulD    & 0s                   & 6,048s         & 319s           & 125.93M         & 1.7792\$       \\
% ChatGPT  &                      &               & 1,263s          & 175,000.00M      & 1.9200\$        \\
% \bottomrule
% \end{tabular}}
% \label{tab:rq7}
% \end{table}

\begin{table}[htbp]
  \centering
    \caption{Time and economic costs of the models}
    \resizebox{\linewidth}{!}{
% \begin{tabular}{lrrrrc}
%          \toprule
%          \textbf{Model} & \textbf{Preprocess Data Time} & \textbf{Training Time} & \textbf{Inference Time} & \textbf{Model Parameter} & \textbf{Cost} \\

\begin{tabular}{lrrrrl}
\toprule
\multirow{2}{*}{\textbf{Model}} & \multicolumn{3}{c}{\textbf{Time}} & \multirow{2}{*}{\textbf{Parameter}} & 
\multirow{2}{*}{\textbf{Cost}} \\
\cmidrule{2-4}  & \textbf{Pre-processing}   & \textbf{Training}  & \textbf{Interring}  &                           &                       \\
         
\midrule
Devign   & 7,103s                & 2,836s         & 101s           & 0.97M           & 2.8056\$       \\
Reveal   & 7,103s                & 5,220s         & 148s           & 1.09M           & 4.4378\$       \\
IVDetect & 3,563s                & 1,3602s        & 916s           & 1.01M           & 4.8821\$       \\
LineVul  & 0s                   & 13,274s        & 322s           & 124.65M         & 6.1712\$       \\
SVulD    & 0s                   & 6,048s         & 319s           & 125.93M         & 1.7792\$       \\
ChatGPT  &                      &               & 1,263s          & 175,000.00M      & 0.6385\$        \\
\bottomrule
\end{tabular}}
\label{tab:rq7}
\end{table}

\noindent
\textbf{Results}.
% \cn{analyze obtained results}
% \xxd
{
The results are summarized in Table~\ref{tab:rq7}. Based on the results, we can obtain the following findings:
(1) 
Graph-based models require a significant time cost for data preprocessing, sometimes exceeding the time needed for model training.
% IVDetect(https://arxiv.org/pdf/2106.10478.pdf)由于其复杂的模型结构，代码经过了复杂的处理，输入到TreeLSTM,GLOVE,GRU。才输入到GNN层，并在后续特征向量使用了Pooling等层，复杂的模型结构，可能是训练时间很大的原因。
IVDetect has the longest training time among graph-based models due to its complex model structure, where the input goes through multiple layers, such as TreeLSTM, GloVe, GNN, and the pooling layer. 
% Reveal是个两阶段模型，首先利用GNN训练获得代码的特征表示，然后再输入到一个表示模型当中继续训练。
% 且Devign和Reveal所使用的GNN模型较为简单，使用了单层的GatedGraphConv，相比一个阶段直接训练的Devign来说，它有着最短的训练时间。
In contrast, the model architectures of Devign and Reveal are relatively simpler. In particular, Devign trains the fastest because it only adopts a single-layer GatedGraphConv.
(2) 
% LineVul 和 SVulD使用了BERT模型，具有12层transformer层，相比于其它图模型有着更多的模型参数。
Sequence-based models, especially ChatGPT, are more complex in structures with larger amounts of model parameters, which explains their longer inference time. 
However, sequence-based models also have advantages: they require zero data preprocessing time; LineVul and SVulD are comparable to graph-based models in training time.
% LineVul 和 SVulD有着相同的结构，虽然epoch设置的一样，均为20，但LineVul因为官方源码没有实现early_stop机制训练的时间更多
Though LineVul and SVulD have similar model structures (i.e., 12 transformer layers) and we set the epoch equally as 20, LineVul requires more training time because its official implementation not adopting an early stopping mechanism.
(3)
% 经济角度
Among all the models, ChatGPT is the most economical option with a cost of only 0.6385\$, which shows its potential for practical usage.
}
% \chao{what is the cost? for each function? or the whole testing dataset?}
% chatgpt 推理的cost
% 其它深度学习模型：数据预处理，训练，推理总共的cost

% \vspace{-0.2cm}
\intuition{
\textbf{Finding 7}:
Graph-based models need large amounts of time for data preprocessing, but they typically train and infer fast.
In contrast, sequence-based models do not involve data preprocessing, with a comparable training time and longer inference time.
Overall, ChatGPT is the most economical solution.
}
% \vspace{-0.3cm}

%% file: sections/threats.tex
% \vspace{-0.3cm}
\section{Threats to Validity}

\textbf{Internal Validity} arises from two aspects. 
The first one is about the uncertainty of LLM's output.
Previous work has verified that LLMs are sensitive to prompts, such as the number and quality of selected examples in-context learning and chain-of-thoughts, and natural language instruction. 
To alleviate this threat, we explore the performance of different example strategies in RQ1 and use fixed instructions and random seeds to ensure the generated content is relatively consistent.
% grid search find best temperature
In addition,  ChatGPT is a closed-source LLM, which poses a threat to reproducibility, so the results we report may relate to a specific version of ChatGPT (i.e., GPT-3.5 Turbo).
Another potential threat is the implementation of a graph-based vulnerability detection model. 
To mitigate this threat, we leverage the open-source implementations provided by previous works. 
In cases where the code is unavailable, we employ paired programming to ensure a close replication of the performance reported in the original paper. 
Furthermore, we strictly adhere to the hyperparameters  reported in the original papers.

% \sly{
% \textbf{Internal Validity} concerns whether our experiments demonstrate causality. 
% The first threat relates to the LLM result determinacy. Previous work has found that LLMs are sensitive to prompts, such as the selected examples in-context learning and chain-of-thoughts, the number of examples and natural language instruction. To alleviate this threat, We explored the performance of different example strategies in RQ1 and used fixed instructions and random seeds to ensure that the generated prompts remained consistent.
% % grid search find best temperature
% In addition, the ChatGPT used in experiments is a closed-source LLM, which poses a threat to reproducibility, so the results we report may relate to a specific version of ChatGPT.
% Another potential internal threat is the implementation of a graph-based vulnerability detection model. 
% To mitigate this threat, We leverage the open-source implementations provided by previous works. 
% In cases where the code is unavailable, we employ paired programming to ensure a close replication of the performance reported in the original paper. 
% Furthermore, we strictly adhere to the hyperparameters reported in the original papers.
% }

\textbf{External Validity} concerns the generalization of our report results.
The first threat comes from the fact that we focus on vulnerability detection in C and C++ languages, many 
 disclosed vulnerabilities in other popular languages (e.g., Java or Python) are not considered in this study.
% Secondly, we do not investigate other types of LLMs (e.g., StarCoder and Codex) since ChatGPT has been verified as the most powerful LLM. 
% that we have left for future research due to budget limitations.
Another threat is the impact of dataset selection.
To mitigate this threat, we have created the \datasetname dataset to cover most of the C/C++ vulnerabilities recorded in the NVD database since 2003, which is the largest function-level vulnerability dataset, ensuring that the evaluation results are representative and convincing.

% \sly{
% \textbf{External Validity} concerns whether the results presented would generalize. 
% The first external validity comes from the fact that in this paper we focuses on vulnerability detection in C and C++ languages, there are a number of other popular languages such as Java, Python, where vulnerability detection is not considered. 
% Secondly, there are other LLMs such as StarCoder and Codex that we have left for future research due to budget limitations.
% Another is the impact of dataset selection, where existing datasets that used in prior works may not reflect the characteristics of real-world vulnerabilities. To mitigate this threat, we have created the Vul4C dataset to cover most of the C/C++ vulnerabilities recorded in the NVD database since 2003, which is the largest known function-level vulnerability dataset, so that the evaluation results should be representative and convincing.
% }

%% file: sections/related_work.tex
\section{Related Work}

Vulnerability detection (VD) has attracted much attention and many learning-based approaches have been proposed to automatically learn the vulnerability patterns from historical data~\cite{yamaguchi2014modeling,li2018vuldeepecker,zhou2019devign,li2021vuldeelocator,duan2019vulsniper,lin2017poster,chakraborty2021deep,li2021sysevr}.
These methods can be further divided into complex graph-based ones~\cite{yamaguchi2014modeling,zhou2019devign,cheng2021deepwukong,wang2020combining,cao2022mvd,hin2022linevd} and sequence-based ones~\cite{dam2017automatic,russell2018automated,fu2022linevul,ni2023distinguishing}, and have become state-of-the-art.

Recently, a few works have conducted empirical studies on  these learning-based vulnerability detection models.
Chakaborthy et al.~\cite{chakraborty2021deep} investigated the issues of synthetic datasets, data duplication, and data imbalance by studying four deep learning models and then improved their model design based on their findings.
Tang et al.~\cite{tang2020comparative} surveyed two models to investigate the best methods among neural network architectures, vector representation methods, and symbolization methods.
% Mazuera-Rozo et al. [28] evaluated 1 shallow and 2 deep models on binary classification and bug type (nonbinary) classification. 
% After we completed our study, we found two related empirical studies. 
Lin et al.~\cite{lin2021deep} construct dataset including nine software projects to evaluate six neural network models' vulnerability detection ability and their generalization.
Meanwhile, Ban et al.~\cite{ban2019performance} evaluated six learning based models in a cross-project setting considering three software projects.
Steenhoek et al.~\cite{steenhoek2023empirical} also conduct an empirical study on deep learning based vulnerability detection models with the consideration of three dimensions (i.e., model capabilities, training data, and model interpretation).

Different from these works, our work extensively studies  the characteristics of learning-based VD approaches in the era of large pre-trained language models, especially focusing on ChatGPT's remarkable ability by considering five dimensions.
To the best of our knowledge, our work is the first attempt to characterize the ChatGPT's ability on VD, the ease of use of models, the model economy, and types of vulnerability that models are skilled in.

%% file: sections/conclusion.tex
% \vspace{-0.3cm}
\section{Conclusion}

% 我们对基于学习的漏洞检测模型进行了深入的研究。我们在统一的数据集Vul4C上进行评估，
% 实验结果表明，基于sequence的模型性能好于基于graph的模型，ChatGPT还不能够很好的胜任漏洞检测任务。
% 基于graph和基于sequenc的方法都主要关注于两种语句类型：Function Call和Field Expression。
% 即使输入变化在语义上是相等的，所有模型都表现出性能下降的趋势，但基于sequence的模型性能下降很少。
% 基于graph的模型需要完整的输入程序用来构造图，相比来说基于sequence的模型更容易部署。
% 基于graph的模型通常需要大量时间进行数据的预处理，但是所需的训练和推理时间比sequence模型花费更少。
% 我们希望安全分析人员通过本篇论文对模型的深入研究，了解现有漏洞检测模型的现状，并根据不同的使用场景选择适合的漏洞检测模型。
% todo

This paper aims to comprehensively investigate the capabilities of graph-based and sequence-based learning-based models for vulnerability detection as well as their impacts.
To achieve that, we first build a large-scale vulnerability dataset and then conduct several experiments focusing on five dimensions: 
\textit{model capabilities}, \textit{model interpretation}, \textit{model stability}, \textit{ease of use of model}, and \textit{model economy}.
The results indicate the priority of sequence-based models and the limited abilities of both LLM (ChatGPT) and graph-based models.
We also investigate the performance of learning-based models on types of vulnerability and find that both sequence-based and graph-based models are skilled in ``Input Validation'', while graph-based models are skilled at another two: ``API Abuse'' and ``Security Feature''.
We also find that all learning-based models perform inconsistently.
Finally, we conclude the pre-processing and requirements for easy usage of models and obtain vital information for economically and safely practical usage of these models.

%% file: main.bbl

\begin{thebibliography}{63}

%%% the Web address.

\ifx \showCODEN    \undefined \def \showCODEN     #1{\unskip}     \fi
\ifx \showDOI      \undefined \def \showDOI       #1{#1}\fi
\ifx \showISBNx    \undefined \def \showISBNx     #1{\unskip}     \fi
\ifx \showISBNxiii \undefined \def \showISBNxiii  #1{\unskip}     \fi
\ifx \showISSN     \undefined \def \showISSN      #1{\unskip}     \fi
\ifx \showLCCN     \undefined \def \showLCCN      #1{\unskip}     \fi
\ifx \shownote     \undefined \def \shownote      #1{#1}          \fi
\ifx \showarticletitle \undefined \def \showarticletitle #1{#1}   \fi
\ifx \showURL      \undefined \def \showURL       {\relax}        \fi
% invisible to TeX
\providecommand\bibfield[2]{#2}
\providecommand\bibinfo[2]{#2}
\providecommand\natexlab[1]{#1}
\providecommand\showeprint[2][]{arXiv:#2}

\bibitem[sar(2018)]%
        {sard}
 \bibinfo{year}{2018}\natexlab{}.
\newblock \bibinfo{title}{Software assurance reference dataset (SARD)}.
\newblock \bibinfo{howpublished}{\url{https://samate.nist.gov/SARD/}}.
\newblock


\bibitem[ope(2022)]%
        {openaichatgpt}
 \bibinfo{year}{2022}\natexlab{}.
\newblock \bibinfo{title}{Chatgpt: Optimizing language models for dialogue}.
\newblock
\newblock
\urldef\tempurl%
\url{https://chat.openai.com}
\showURL{%
\tempurl}


\bibitem[aws(2024)]%
        {aws_g5}
 \bibinfo{year}{2024}\natexlab{}.
\newblock \bibinfo{title}{AWS g5 instance}.
\newblock
  \bibinfo{howpublished}{\url{https://aws.amazon.com/cn/ec2/instance-types/g5/}}.
\newblock


\bibitem[CVE(2024)]%
        {CVE-2022-47519}
 \bibinfo{year}{2024}\natexlab{}.
\newblock \bibinfo{title}{CVE-2022-47519}.
\newblock
  \bibinfo{howpublished}{\url{https://github.com/torvalds/linux/commit/051ae669e4505abbe05165bebf6be7922de11f41}}.
\newblock


\bibitem[hug(2024)]%
        {huggingface}
 \bibinfo{year}{2024}\natexlab{}.
\newblock \bibinfo{title}{Hugging Face}.
\newblock
\newblock
\urldef\tempurl%
\url{https://huggingface.co}
\showURL{%
\tempurl}


\bibitem[ope(2024)]%
        {openai-pricing}
 \bibinfo{year}{2024}\natexlab{}.
\newblock \bibinfo{title}{OpenAI Pricing}.
\newblock \bibinfo{howpublished}{\url{https://openai.com/pricing}}.
\newblock


\bibitem[rep(2024)]%
        {replication}
 \bibinfo{year}{2024}\natexlab{}.
\newblock \bibinfo{title}{Replication}.
\newblock
\newblock
\urldef\tempurl%
\url{https://figshare.com/s/bde8e41890e8179fbe5f}
\showURL{%
\tempurl}


\bibitem[tre(2024)]%
        {tree-sitter}
 \bibinfo{year}{2024}\natexlab{}.
\newblock \bibinfo{title}{Tree-sitter}.
\newblock
  \bibinfo{howpublished}{\url{https://github.com/tree-sitter/tree-sitter}}.
\newblock


\bibitem[Ban et~al\mbox{.}(2019)]%
        {ban2019performance}
\bibfield{author}{\bibinfo{person}{Xinbo Ban}, \bibinfo{person}{Shigang Liu},
  \bibinfo{person}{Chao Chen}, {and} \bibinfo{person}{Caslon Chua}.}
  \bibinfo{year}{2019}\natexlab{}.
\newblock \showarticletitle{A performance evaluation of deep-learnt features
  for software vulnerability detection}.
\newblock \bibinfo{journal}{\emph{Concurrency and Computation: Practice and
  Experience}} \bibinfo{volume}{31}, \bibinfo{number}{19}
  (\bibinfo{year}{2019}), \bibinfo{pages}{e5103}.
\newblock


\bibitem[Brown et~al\mbox{.}(2020)]%
        {brown2020language}
\bibfield{author}{\bibinfo{person}{Tom Brown}, \bibinfo{person}{Benjamin Mann},
  \bibinfo{person}{Nick Ryder}, \bibinfo{person}{Melanie Subbiah},
  \bibinfo{person}{Jared~D Kaplan}, \bibinfo{person}{Prafulla Dhariwal},
  \bibinfo{person}{Arvind Neelakantan}, \bibinfo{person}{Pranav Shyam},
  \bibinfo{person}{Girish Sastry}, \bibinfo{person}{Amanda Askell},
  {et~al\mbox{.}}} \bibinfo{year}{2020}\natexlab{}.
\newblock \showarticletitle{Language models are few-shot learners}.
\newblock \bibinfo{journal}{\emph{Advances in neural information processing
  systems}}  \bibinfo{volume}{33} (\bibinfo{year}{2020}),
  \bibinfo{pages}{1877--1901}.
\newblock


\bibitem[Cao et~al\mbox{.}(2022)]%
        {cao2022mvd}
\bibfield{author}{\bibinfo{person}{Sicong Cao}, \bibinfo{person}{Xiaobing Sun},
  \bibinfo{person}{Lili Bo}, \bibinfo{person}{Rongxin Wu}, \bibinfo{person}{Bin
  Li}, {and} \bibinfo{person}{Chuanqi Tao}.} \bibinfo{year}{2022}\natexlab{}.
\newblock \showarticletitle{MVD: Memory-Related Vulnerability Detection Based
  on Flow-Sensitive Graph Neural Networks}.
\newblock \bibinfo{journal}{\emph{arXiv preprint arXiv:2203.02660}}
  (\bibinfo{year}{2022}).
\newblock


\bibitem[Chakraborty et~al\mbox{.}(2021)]%
        {chakraborty2021deep}
\bibfield{author}{\bibinfo{person}{Saikat Chakraborty}, \bibinfo{person}{Rahul
  Krishna}, \bibinfo{person}{Yangruibo Ding}, {and} \bibinfo{person}{Baishakhi
  Ray}.} \bibinfo{year}{2021}\natexlab{}.
\newblock \showarticletitle{Deep learning based vulnerability detection: Are we
  there yet}.
\newblock \bibinfo{journal}{\emph{IEEE Transactions on Software Engineering}}
  (\bibinfo{year}{2021}).
\newblock


\bibitem[Cheng et~al\mbox{.}(2021)]%
        {cheng2021deepwukong}
\bibfield{author}{\bibinfo{person}{Xiao Cheng}, \bibinfo{person}{Haoyu Wang},
  \bibinfo{person}{Jiayi Hua}, \bibinfo{person}{Guoai Xu}, {and}
  \bibinfo{person}{Yulei Sui}.} \bibinfo{year}{2021}\natexlab{}.
\newblock \showarticletitle{Deepwukong: Statically detecting software
  vulnerabilities using deep graph neural network}.
\newblock \bibinfo{journal}{\emph{ACM Transactions on Software Engineering and
  Methodology (TOSEM)}} \bibinfo{volume}{30}, \bibinfo{number}{3}
  (\bibinfo{year}{2021}), \bibinfo{pages}{1--33}.
\newblock


\bibitem[Croft et~al\mbox{.}(2023)]%
        {croft2023data}
\bibfield{author}{\bibinfo{person}{Roland Croft}, \bibinfo{person}{M~Ali
  Babar}, {and} \bibinfo{person}{M~Mehdi Kholoosi}.}
  \bibinfo{year}{2023}\natexlab{}.
\newblock \showarticletitle{Data quality for software vulnerability datasets}.
  In \bibinfo{booktitle}{\emph{2023 IEEE/ACM 45th International Conference on
  Software Engineering (ICSE)}}. IEEE, \bibinfo{pages}{121--133}.
\newblock


\bibitem[Dam et~al\mbox{.}(2017)]%
        {dam2017automatic}
\bibfield{author}{\bibinfo{person}{Hoa~Khanh Dam}, \bibinfo{person}{Truyen
  Tran}, \bibinfo{person}{Trang Pham}, \bibinfo{person}{Shien~Wee Ng},
  \bibinfo{person}{John Grundy}, {and} \bibinfo{person}{Aditya Ghose}.}
  \bibinfo{year}{2017}\natexlab{}.
\newblock \showarticletitle{Automatic feature learning for vulnerability
  prediction}.
\newblock \bibinfo{journal}{\emph{arXiv preprint arXiv:1708.02368}}
  (\bibinfo{year}{2017}).
\newblock


\bibitem[Duan et~al\mbox{.}(2019)]%
        {duan2019vulsniper}
\bibfield{author}{\bibinfo{person}{Xu Duan}, \bibinfo{person}{Jingzheng Wu},
  \bibinfo{person}{Shouling Ji}, \bibinfo{person}{Zhiqing Rui},
  \bibinfo{person}{Tianyue Luo}, \bibinfo{person}{Mutian Yang}, {and}
  \bibinfo{person}{Yanjun Wu}.} \bibinfo{year}{2019}\natexlab{}.
\newblock \showarticletitle{VulSniper: Focus Your Attention to Shoot
  Fine-Grained Vulnerabilities.}. In \bibinfo{booktitle}{\emph{IJCAI}}.
  \bibinfo{pages}{4665--4671}.
\newblock


\bibitem[Fan et~al\mbox{.}(2020)]%
        {fan2020ac}
\bibfield{author}{\bibinfo{person}{Jiahao Fan}, \bibinfo{person}{Yi Li},
  \bibinfo{person}{Shaohua Wang}, {and} \bibinfo{person}{Tien~N Nguyen}.}
  \bibinfo{year}{2020}\natexlab{}.
\newblock \showarticletitle{A C/C++ code vulnerability dataset with code
  changes and CVE summaries}. In \bibinfo{booktitle}{\emph{Proceedings of the
  17th International Conference on Mining Software Repositories}}.
  \bibinfo{pages}{508--512}.
\newblock


\bibitem[Feng et~al\mbox{.}(2020b)]%
        {feng2020codebert}
\bibfield{author}{\bibinfo{person}{Zhangyin Feng}, \bibinfo{person}{Daya Guo},
  \bibinfo{person}{Duyu Tang}, \bibinfo{person}{Nan Duan},
  \bibinfo{person}{Xiaocheng Feng}, \bibinfo{person}{Ming Gong},
  \bibinfo{person}{Linjun Shou}, \bibinfo{person}{Bing Qin},
  \bibinfo{person}{Ting Liu}, \bibinfo{person}{Daxin Jiang}, {et~al\mbox{.}}}
  \bibinfo{year}{2020}\natexlab{b}.
\newblock \showarticletitle{Codebert: A pre-trained model for programming and
  natural languages}.
\newblock \bibinfo{journal}{\emph{arXiv preprint arXiv:2002.08155}}
  (\bibinfo{year}{2020}).
\newblock


\bibitem[Feng et~al\mbox{.}(2020a)]%
        {fengetal2020codebert}
\bibfield{author}{\bibinfo{person}{Zhangyin Feng}, \bibinfo{person}{Daya Guo},
  \bibinfo{person}{Duyu Tang}, \bibinfo{person}{Nan Duan},
  \bibinfo{person}{Xiaocheng Feng}, \bibinfo{person}{Ming Gong},
  \bibinfo{person}{Linjun Shou}, \bibinfo{person}{Bing Qin},
  \bibinfo{person}{Ting Liu}, \bibinfo{person}{Daxin Jiang}, {and}
  \bibinfo{person}{Ming Zhou}.} \bibinfo{year}{2020}\natexlab{a}.
\newblock \showarticletitle{{C}ode{BERT}: A Pre-Trained Model for Programming
  and Natural Languages}. In \bibinfo{booktitle}{\emph{Findings of the
  Association for Computational Linguistics: EMNLP 2020}}.
  \bibinfo{publisher}{Association for Computational Linguistics},
  \bibinfo{address}{Online}, \bibinfo{pages}{1536--1547}.
\newblock
\urldef\tempurl%
\url{https://doi.org/10.18653/v1/2020.findings-emnlp.139}
\showDOI{\tempurl}


\bibitem[Fu and Tantithamthavorn(2022)]%
        {fu2022linevul}
\bibfield{author}{\bibinfo{person}{Michael Fu} {and} \bibinfo{person}{Chakkrit
  Tantithamthavorn}.} \bibinfo{year}{2022}\natexlab{}.
\newblock \showarticletitle{LineVul: A Transformer-based Line-Level
  Vulnerability Prediction}.
\newblock  (\bibinfo{year}{2022}).
\newblock


\bibitem[Guo et~al\mbox{.}(2022)]%
        {guo2022unixcoder}
\bibfield{author}{\bibinfo{person}{Daya Guo}, \bibinfo{person}{Shuai Lu},
  \bibinfo{person}{Nan Duan}, \bibinfo{person}{Yanlin Wang},
  \bibinfo{person}{Ming Zhou}, {and} \bibinfo{person}{Jian Yin}.}
  \bibinfo{year}{2022}\natexlab{}.
\newblock \showarticletitle{UniXcoder: Unified Cross-Modal Pre-training for
  Code Representation}.
\newblock \bibinfo{journal}{\emph{arXiv preprint arXiv:2203.03850}}
  (\bibinfo{year}{2022}).
\newblock


\bibitem[Hanif and Maffeis(2022)]%
        {hanif2022vulberta}
\bibfield{author}{\bibinfo{person}{Hazim Hanif} {and} \bibinfo{person}{Sergio
  Maffeis}.} \bibinfo{year}{2022}\natexlab{}.
\newblock \showarticletitle{Vulberta: Simplified source code pre-training for
  vulnerability detection}. In \bibinfo{booktitle}{\emph{2022 International
  joint conference on neural networks (IJCNN)}}. IEEE, \bibinfo{pages}{1--8}.
\newblock


\bibitem[Hin et~al\mbox{.}(2022)]%
        {hin2022linevd}
\bibfield{author}{\bibinfo{person}{David Hin}, \bibinfo{person}{Andrey Kan},
  \bibinfo{person}{Huaming Chen}, {and} \bibinfo{person}{M~Ali Babar}.}
  \bibinfo{year}{2022}\natexlab{}.
\newblock \showarticletitle{LineVD: Statement-level Vulnerability Detection
  using Graph Neural Networks}.
\newblock \bibinfo{journal}{\emph{arXiv preprint arXiv:2203.05181}}
  (\bibinfo{year}{2022}).
\newblock


\bibitem[Kojima et~al\mbox{.}(2022)]%
        {kojima2022large}
\bibfield{author}{\bibinfo{person}{Takeshi Kojima},
  \bibinfo{person}{Shixiang~Shane Gu}, \bibinfo{person}{Machel Reid},
  \bibinfo{person}{Yutaka Matsuo}, {and} \bibinfo{person}{Yusuke Iwasawa}.}
  \bibinfo{year}{2022}\natexlab{}.
\newblock \showarticletitle{Large language models are zero-shot reasoners}.
\newblock \bibinfo{journal}{\emph{Advances in neural information processing
  systems}}  \bibinfo{volume}{35} (\bibinfo{year}{2022}),
  \bibinfo{pages}{22199--22213}.
\newblock


\bibitem[Li et~al\mbox{.}(2017)]%
        {li2017large}
\bibfield{author}{\bibinfo{person}{Bo Li}, \bibinfo{person}{Kevin Roundy},
  \bibinfo{person}{Chris Gates}, {and} \bibinfo{person}{Yevgeniy Vorobeychik}.}
  \bibinfo{year}{2017}\natexlab{}.
\newblock \showarticletitle{Large-scale identification of malicious singleton
  files}. In \bibinfo{booktitle}{\emph{Proceedings of the Seventh ACM on
  Conference on Data and Application Security and Privacy}}.
  \bibinfo{pages}{227--238}.
\newblock


\bibitem[Li et~al\mbox{.}(2021a)]%
        {li2021vulnerability}
\bibfield{author}{\bibinfo{person}{Yi Li}, \bibinfo{person}{Shaohua Wang},
  {and} \bibinfo{person}{Tien~N Nguyen}.} \bibinfo{year}{2021}\natexlab{a}.
\newblock \showarticletitle{Vulnerability detection with fine-grained
  interpretations}. In \bibinfo{booktitle}{\emph{Proceedings of the 29th ACM
  Joint Meeting on European Software Engineering Conference and Symposium on
  the Foundations of Software Engineering}}. \bibinfo{pages}{292--303}.
\newblock


\bibitem[Li et~al\mbox{.}(2021b)]%
        {li2021vuldeelocator}
\bibfield{author}{\bibinfo{person}{Zhen Li}, \bibinfo{person}{Deqing Zou},
  \bibinfo{person}{Shouhuai Xu}, \bibinfo{person}{Zhaoxuan Chen},
  \bibinfo{person}{Yawei Zhu}, {and} \bibinfo{person}{Hai Jin}.}
  \bibinfo{year}{2021}\natexlab{b}.
\newblock \showarticletitle{Vuldeelocator: a deep learning-based fine-grained
  vulnerability detector}.
\newblock \bibinfo{journal}{\emph{IEEE Transactions on Dependable and Secure
  Computing}} (\bibinfo{year}{2021}).
\newblock


\bibitem[Li et~al\mbox{.}(2021c)]%
        {li2021sysevr}
\bibfield{author}{\bibinfo{person}{Zhen Li}, \bibinfo{person}{Deqing Zou},
  \bibinfo{person}{Shouhuai Xu}, \bibinfo{person}{Hai Jin},
  \bibinfo{person}{Yawei Zhu}, {and} \bibinfo{person}{Zhaoxuan Chen}.}
  \bibinfo{year}{2021}\natexlab{c}.
\newblock \showarticletitle{Sysevr: A framework for using deep learning to
  detect software vulnerabilities}.
\newblock \bibinfo{journal}{\emph{IEEE Transactions on Dependable and Secure
  Computing}} (\bibinfo{year}{2021}).
\newblock


\bibitem[Li et~al\mbox{.}(2018)]%
        {li2018vuldeepecker}
\bibfield{author}{\bibinfo{person}{Zhen Li}, \bibinfo{person}{Deqing Zou},
  \bibinfo{person}{Shouhuai Xu}, \bibinfo{person}{Xinyu Ou},
  \bibinfo{person}{Hai Jin}, \bibinfo{person}{Sujuan Wang},
  \bibinfo{person}{Zhijun Deng}, {and} \bibinfo{person}{Yuyi Zhong}.}
  \bibinfo{year}{2018}\natexlab{}.
\newblock \showarticletitle{Vuldeepecker: A deep learning-based system for
  vulnerability detection}. In \bibinfo{booktitle}{\emph{Proceedings of the
  25th Annual Network and Distributed System Security Symposium}}.
\newblock


\bibitem[Lin et~al\mbox{.}(2021)]%
        {lin2021deep}
\bibfield{author}{\bibinfo{person}{Guanjun Lin}, \bibinfo{person}{Wei Xiao},
  \bibinfo{person}{Leo~Yu Zhang}, \bibinfo{person}{Shang Gao},
  \bibinfo{person}{Yonghang Tai}, {and} \bibinfo{person}{Jun Zhang}.}
  \bibinfo{year}{2021}\natexlab{}.
\newblock \showarticletitle{Deep neural-based vulnerability discovery
  demystified: data, model and performance}.
\newblock \bibinfo{journal}{\emph{Neural Computing and Applications}}
  \bibinfo{volume}{33}, \bibinfo{number}{20} (\bibinfo{year}{2021}),
  \bibinfo{pages}{13287--13300}.
\newblock


\bibitem[Lin et~al\mbox{.}(2017)]%
        {lin2017poster}
\bibfield{author}{\bibinfo{person}{Guanjun Lin}, \bibinfo{person}{Jun Zhang},
  \bibinfo{person}{Wei Luo}, \bibinfo{person}{Lei Pan}, {and}
  \bibinfo{person}{Yang Xiang}.} \bibinfo{year}{2017}\natexlab{}.
\newblock \showarticletitle{POSTER: Vulnerability discovery with function
  representation learning from unlabeled projects}. In
  \bibinfo{booktitle}{\emph{Proceedings of the 2017 ACM SIGSAC Conference on
  Computer and Communications Security}}. \bibinfo{pages}{2539--2541}.
\newblock


\bibitem[Liu et~al\mbox{.}(2021)]%
        {liu2021makes}
\bibfield{author}{\bibinfo{person}{Jiachang Liu}, \bibinfo{person}{Dinghan
  Shen}, \bibinfo{person}{Yizhe Zhang}, \bibinfo{person}{Bill Dolan},
  \bibinfo{person}{Lawrence Carin}, {and} \bibinfo{person}{Weizhu Chen}.}
  \bibinfo{year}{2021}\natexlab{}.
\newblock \showarticletitle{What Makes Good In-Context Examples for GPT-$3 $?}
\newblock \bibinfo{journal}{\emph{arXiv preprint arXiv:2101.06804}}
  (\bibinfo{year}{2021}).
\newblock


\bibitem[Liu et~al\mbox{.}(2023)]%
        {liu2023pre}
\bibfield{author}{\bibinfo{person}{Pengfei Liu}, \bibinfo{person}{Weizhe Yuan},
  \bibinfo{person}{Jinlan Fu}, \bibinfo{person}{Zhengbao Jiang},
  \bibinfo{person}{Hiroaki Hayashi}, {and} \bibinfo{person}{Graham Neubig}.}
  \bibinfo{year}{2023}\natexlab{}.
\newblock \showarticletitle{Pre-train, prompt, and predict: A systematic survey
  of prompting methods in natural language processing}.
\newblock \bibinfo{journal}{\emph{Comput. Surveys}} \bibinfo{volume}{55},
  \bibinfo{number}{9} (\bibinfo{year}{2023}), \bibinfo{pages}{1--35}.
\newblock


\bibitem[Lu et~al\mbox{.}(2021)]%
        {lu2021fantastically}
\bibfield{author}{\bibinfo{person}{Yao Lu}, \bibinfo{person}{Max Bartolo},
  \bibinfo{person}{Alastair Moore}, \bibinfo{person}{Sebastian Riedel}, {and}
  \bibinfo{person}{Pontus Stenetorp}.} \bibinfo{year}{2021}\natexlab{}.
\newblock \showarticletitle{Fantastically ordered prompts and where to find
  them: Overcoming few-shot prompt order sensitivity}.
\newblock \bibinfo{journal}{\emph{arXiv preprint arXiv:2104.08786}}
  (\bibinfo{year}{2021}).
\newblock


\bibitem[MacQueen et~al\mbox{.}(1967)]%
        {macqueen1967some}
\bibfield{author}{\bibinfo{person}{James MacQueen} {et~al\mbox{.}}}
  \bibinfo{year}{1967}\natexlab{}.
\newblock \showarticletitle{Some methods for classification and analysis of
  multivariate observations}. In \bibinfo{booktitle}{\emph{Proceedings of the
  fifth Berkeley symposium on mathematical statistics and probability}},
  Vol.~\bibinfo{volume}{1}. Oakland, CA, USA, \bibinfo{pages}{281--297}.
\newblock


\bibitem[Maiorca and Biggio(2019)]%
        {maiorca2019digital}
\bibfield{author}{\bibinfo{person}{Davide Maiorca} {and}
  \bibinfo{person}{Battista Biggio}.} \bibinfo{year}{2019}\natexlab{}.
\newblock \showarticletitle{Digital investigation of pdf files: Unveiling
  traces of embedded malware}.
\newblock \bibinfo{journal}{\emph{IEEE Security \& Privacy}}
  \bibinfo{volume}{17}, \bibinfo{number}{1} (\bibinfo{year}{2019}),
  \bibinfo{pages}{63--71}.
\newblock


\bibitem[Min et~al\mbox{.}(2022)]%
        {min-etal-2022-rethinking}
\bibfield{author}{\bibinfo{person}{Sewon Min}, \bibinfo{person}{Xinxi Lyu},
  \bibinfo{person}{Ari Holtzman}, \bibinfo{person}{Mikel Artetxe},
  \bibinfo{person}{Mike Lewis}, \bibinfo{person}{Hannaneh Hajishirzi}, {and}
  \bibinfo{person}{Luke Zettlemoyer}.} \bibinfo{year}{2022}\natexlab{}.
\newblock \showarticletitle{Rethinking the Role of Demonstrations: What Makes
  In-Context Learning Work?}. In \bibinfo{booktitle}{\emph{Proceedings of the
  2022 Conference on Empirical Methods in Natural Language Processing}}.
  \bibinfo{publisher}{Association for Computational Linguistics},
  \bibinfo{address}{Abu Dhabi, United Arab Emirates},
  \bibinfo{pages}{11048--11064}.
\newblock
\urldef\tempurl%
\url{https://doi.org/10.18653/v1/2022.emnlp-main.759}
\showDOI{\tempurl}


\bibitem[Ni et~al\mbox{.}(2024)]%
        {ni2024megavul}
\bibfield{author}{\bibinfo{person}{Chao Ni}, \bibinfo{person}{Liyu Shen},
  \bibinfo{person}{Xiaohu Yang}, \bibinfo{person}{Yan Zhu}, {and}
  \bibinfo{person}{Shaohua Wang}.} \bibinfo{year}{2024}\natexlab{}.
\newblock \showarticletitle{MegaVul: A C/C++ Vulnerability Dataset with
  Comprehensive Code Representation}. In \bibinfo{booktitle}{\emph{Proceedings
  of 21th International Conference on Mining Software Repositories (MSR)}}.
\newblock


\bibitem[Ni et~al\mbox{.}(2022a)]%
        {ni2022best}
\bibfield{author}{\bibinfo{person}{Chao Ni}, \bibinfo{person}{Wei Wang},
  \bibinfo{person}{Kaiwen Yang}, \bibinfo{person}{Xin Xia},
  \bibinfo{person}{Kui Liu}, {and} \bibinfo{person}{David Lo}.}
  \bibinfo{year}{2022}\natexlab{a}.
\newblock \showarticletitle{{ The Best of Both Worlds: Integrating Semantic
  Features with Expert Features for Defect Prediction and Localization}}. In
  \bibinfo{booktitle}{\emph{Proceedings of the 2022 30th ACM Joint Meeting on
  European Software Engineering Conference and Symposium on the Foundations of
  Software Engineering}}. ACM, \bibinfo{pages}{672--683}.
\newblock


\bibitem[Ni et~al\mbox{.}(2022b)]%
        {ni2022defect}
\bibfield{author}{\bibinfo{person}{Chao Ni}, \bibinfo{person}{Kaiwen Yang},
  \bibinfo{person}{Xin Xia}, \bibinfo{person}{David Lo}, \bibinfo{person}{Xiang
  Chen}, {and} \bibinfo{person}{Xiaohu Yang}.}
  \bibinfo{year}{2022}\natexlab{b}.
\newblock \showarticletitle{Defect Identification, Categorization, and Repair:
  Better Together}.
\newblock \bibinfo{journal}{\emph{arXiv preprint arXiv:2204.04856}}
  (\bibinfo{year}{2022}).
\newblock


\bibitem[Ni et~al\mbox{.}(2023)]%
        {ni2023distinguishing}
\bibfield{author}{\bibinfo{person}{Chao Ni}, \bibinfo{person}{Xin Yin},
  \bibinfo{person}{Kaiwen Yang}, \bibinfo{person}{Dehai Zhao},
  \bibinfo{person}{Zhenchang Xing}, {and} \bibinfo{person}{Xin Xia}.}
  \bibinfo{year}{2023}\natexlab{}.
\newblock \showarticletitle{Distinguishing Look-Alike Innocent and Vulnerable
  Code by Subtle Semantic Representation Learning and Explanation}. In
  \bibinfo{booktitle}{\emph{Proceedings of the 31st ACM Joint European Software
  Engineering Conference and Symposium on the Foundations of Software
  Engineering}}. \bibinfo{pages}{1611--1622}.
\newblock


\bibitem[OpenAI(2022)]%
        {openai2022chatgpt}
\bibfield{author}{\bibinfo{person}{OpenAI}.} \bibinfo{year}{2022}\natexlab{}.
\newblock \bibinfo{title}{ChatGPT: Optimizing Language Models for Dialogue.
  (2022)}.
\newblock \bibinfo{howpublished}{\url{https://openai.com/blog/chatgpt/}}.
\newblock


\bibitem[Paszke et~al\mbox{.}(2019)]%
        {pytorch}
\bibfield{author}{\bibinfo{person}{Adam Paszke}, \bibinfo{person}{Sam Gross},
  \bibinfo{person}{Francisco Massa}, \bibinfo{person}{Adam Lerer},
  \bibinfo{person}{James Bradbury}, \bibinfo{person}{Gregory Chanan},
  \bibinfo{person}{Trevor Killeen}, \bibinfo{person}{Zeming Lin},
  \bibinfo{person}{Natalia Gimelshein}, \bibinfo{person}{Luca Antiga},
  \bibinfo{person}{Alban Desmaison}, \bibinfo{person}{Andreas Kopf},
  \bibinfo{person}{Edward Yang}, \bibinfo{person}{Zachary DeVito},
  \bibinfo{person}{Martin Raison}, \bibinfo{person}{Alykhan Tejani},
  \bibinfo{person}{Sasank Chilamkurthy}, \bibinfo{person}{Benoit Steiner},
  \bibinfo{person}{Lu Fang}, \bibinfo{person}{Junjie Bai}, {and}
  \bibinfo{person}{Soumith Chintala}.} \bibinfo{year}{2019}\natexlab{}.
\newblock \showarticletitle{PyTorch: An Imperative Style, High-Performance Deep
  Learning Library}.
\newblock In \bibinfo{booktitle}{\emph{Advances in Neural Information
  Processing Systems 32}}. \bibinfo{publisher}{Curran Associates, Inc.},
  \bibinfo{pages}{8024--8035}.
\newblock
\urldef\tempurl%
\url{http://papers.neurips.cc/paper/9015-pytorch-an-imperative-style-high-performance-deep-learning-library.pdf}
\showURL{%
\tempurl}


\bibitem[Russell et~al\mbox{.}(2018)]%
        {russell2018automated}
\bibfield{author}{\bibinfo{person}{Rebecca Russell}, \bibinfo{person}{Louis
  Kim}, \bibinfo{person}{Lei Hamilton}, \bibinfo{person}{Tomo Lazovich},
  \bibinfo{person}{Jacob Harer}, \bibinfo{person}{Onur Ozdemir},
  \bibinfo{person}{Paul Ellingwood}, {and} \bibinfo{person}{Marc McConley}.}
  \bibinfo{year}{2018}\natexlab{}.
\newblock \showarticletitle{Automated vulnerability detection in source code
  using deep representation learning}. In \bibinfo{booktitle}{\emph{2018 17th
  IEEE international conference on machine learning and applications (ICMLA)}}.
  IEEE, \bibinfo{pages}{757--762}.
\newblock


\bibitem[Serrano and Smith(2019)]%
        {serrano2019attention}
\bibfield{author}{\bibinfo{person}{Sofia Serrano} {and} \bibinfo{person}{Noah~A
  Smith}.} \bibinfo{year}{2019}\natexlab{}.
\newblock \showarticletitle{Is attention interpretable?}
\newblock \bibinfo{journal}{\emph{arXiv preprint arXiv:1906.03731}}
  (\bibinfo{year}{2019}).
\newblock


\bibitem[Shrikumar et~al\mbox{.}(2019)]%
        {shrikumar2019learning}
\bibfield{author}{\bibinfo{person}{Avanti Shrikumar}, \bibinfo{person}{Peyton
  Greenside}, {and} \bibinfo{person}{Anshul Kundaje}.}
  \bibinfo{year}{2019}\natexlab{}.
\newblock \bibinfo{title}{Learning Important Features Through Propagating
  Activation Differences}.
\newblock
\newblock
\showeprint[arxiv]{1704.02685}~[cs.CV]


\bibitem[Song et~al\mbox{.}(2022)]%
        {song2022hgvul}
\bibfield{author}{\bibinfo{person}{Zihua Song}, \bibinfo{person}{Junfeng Wang},
  \bibinfo{person}{Shengli Liu}, \bibinfo{person}{Zhiyang Fang},
  \bibinfo{person}{Kaiyuan Yang}, {et~al\mbox{.}}}
  \bibinfo{year}{2022}\natexlab{}.
\newblock \showarticletitle{HGVul: A code vulnerability detection method based
  on heterogeneous source-level intermediate representation}.
\newblock \bibinfo{journal}{\emph{Security and Communication Networks}}
  \bibinfo{volume}{2022} (\bibinfo{year}{2022}).
\newblock


\bibitem[Steenhoek et~al\mbox{.}(2023)]%
        {steenhoek2023empirical}
\bibfield{author}{\bibinfo{person}{Benjamin Steenhoek},
  \bibinfo{person}{Md~Mahbubur Rahman}, \bibinfo{person}{Richard Jiles}, {and}
  \bibinfo{person}{Wei Le}.} \bibinfo{year}{2023}\natexlab{}.
\newblock \showarticletitle{An empirical study of deep learning models for
  vulnerability detection}. In \bibinfo{booktitle}{\emph{2023 IEEE/ACM 45th
  International Conference on Software Engineering (ICSE)}}. IEEE,
  \bibinfo{pages}{2237--2248}.
\newblock


\bibitem[Suarez-Tangil et~al\mbox{.}(2017)]%
        {suarez2017droidsieve}
\bibfield{author}{\bibinfo{person}{Guillermo Suarez-Tangil},
  \bibinfo{person}{Santanu~Kumar Dash}, \bibinfo{person}{Mansour Ahmadi},
  \bibinfo{person}{Johannes Kinder}, \bibinfo{person}{Giorgio Giacinto}, {and}
  \bibinfo{person}{Lorenzo Cavallaro}.} \bibinfo{year}{2017}\natexlab{}.
\newblock \showarticletitle{Droidsieve: Fast and accurate classification of
  obfuscated android malware}. In \bibinfo{booktitle}{\emph{Proceedings of the
  seventh ACM on conference on data and application security and privacy}}.
  \bibinfo{pages}{309--320}.
\newblock


\bibitem[Tang et~al\mbox{.}(2020)]%
        {tang2020comparative}
\bibfield{author}{\bibinfo{person}{Gaigai Tang}, \bibinfo{person}{Lianxiao
  Meng}, \bibinfo{person}{Huiqiang Wang}, \bibinfo{person}{Shuangyin Ren},
  \bibinfo{person}{Qiang Wang}, \bibinfo{person}{Lin Yang}, {and}
  \bibinfo{person}{Weipeng Cao}.} \bibinfo{year}{2020}\natexlab{}.
\newblock \showarticletitle{A comparative study of neural network techniques
  for automatic software vulnerability detection}. In
  \bibinfo{booktitle}{\emph{2020 International symposium on theoretical aspects
  of software engineering (TASE)}}. IEEE, \bibinfo{pages}{1--8}.
\newblock


\bibitem[Tsipenyuk et~al\mbox{.}(2005)]%
        {tsipenyuk2005seven}
\bibfield{author}{\bibinfo{person}{Katrina Tsipenyuk}, \bibinfo{person}{Brian
  Chess}, {and} \bibinfo{person}{Gary McGraw}.}
  \bibinfo{year}{2005}\natexlab{}.
\newblock \showarticletitle{Seven pernicious kingdoms: A taxonomy of software
  security errors}.
\newblock \bibinfo{journal}{\emph{IEEE Security \& Privacy}}
  \bibinfo{volume}{3}, \bibinfo{number}{6} (\bibinfo{year}{2005}),
  \bibinfo{pages}{81--84}.
\newblock


\bibitem[van~der Maaten and Hinton(2008)]%
        {JMLR:v9:vandermaaten08a}
\bibfield{author}{\bibinfo{person}{Laurens van~der Maaten} {and}
  \bibinfo{person}{Geoffrey Hinton}.} \bibinfo{year}{2008}\natexlab{}.
\newblock \showarticletitle{Visualizing Data using t-SNE}.
\newblock \bibinfo{journal}{\emph{Journal of Machine Learning Research}}
  \bibinfo{volume}{9}, \bibinfo{number}{86} (\bibinfo{year}{2008}),
  \bibinfo{pages}{2579--2605}.
\newblock
\urldef\tempurl%
\url{http://jmlr.org/papers/v9/vandermaaten08a.html}
\showURL{%
\tempurl}


\bibitem[Vaswani et~al\mbox{.}(2017)]%
        {vaswani2017attention}
\bibfield{author}{\bibinfo{person}{Ashish Vaswani}, \bibinfo{person}{Noam
  Shazeer}, \bibinfo{person}{Niki Parmar}, \bibinfo{person}{Jakob Uszkoreit},
  \bibinfo{person}{Llion Jones}, \bibinfo{person}{Aidan~N Gomez},
  \bibinfo{person}{{\L}ukasz Kaiser}, {and} \bibinfo{person}{Illia
  Polosukhin}.} \bibinfo{year}{2017}\natexlab{}.
\newblock \showarticletitle{Attention is all you need}.
\newblock \bibinfo{journal}{\emph{Advances in neural information processing
  systems}}  \bibinfo{volume}{30} (\bibinfo{year}{2017}).
\newblock


\bibitem[Vaswani et~al\mbox{.}(2023)]%
        {vaswani2023attention}
\bibfield{author}{\bibinfo{person}{Ashish Vaswani}, \bibinfo{person}{Noam
  Shazeer}, \bibinfo{person}{Niki Parmar}, \bibinfo{person}{Jakob Uszkoreit},
  \bibinfo{person}{Llion Jones}, \bibinfo{person}{Aidan~N. Gomez},
  \bibinfo{person}{Lukasz Kaiser}, {and} \bibinfo{person}{Illia Polosukhin}.}
  \bibinfo{year}{2023}\natexlab{}.
\newblock \bibinfo{title}{Attention Is All You Need}.
\newblock
\newblock
\showeprint[arxiv]{1706.03762}~[cs.CL]


\bibitem[Wang et~al\mbox{.}(2020)]%
        {wang2020combining}
\bibfield{author}{\bibinfo{person}{Huanting Wang}, \bibinfo{person}{Guixin Ye},
  \bibinfo{person}{Zhanyong Tang}, \bibinfo{person}{Shin~Hwei Tan},
  \bibinfo{person}{Songfang Huang}, \bibinfo{person}{Dingyi Fang},
  \bibinfo{person}{Yansong Feng}, \bibinfo{person}{Lizhong Bian}, {and}
  \bibinfo{person}{Zheng Wang}.} \bibinfo{year}{2020}\natexlab{}.
\newblock \showarticletitle{Combining graph-based learning with automated data
  collection for code vulnerability detection}.
\newblock \bibinfo{journal}{\emph{IEEE Transactions on Information Forensics
  and Security}}  \bibinfo{volume}{16} (\bibinfo{year}{2020}),
  \bibinfo{pages}{1943--1958}.
\newblock


\bibitem[Wang et~al\mbox{.}(2023)]%
        {wang2023deepvd}
\bibfield{author}{\bibinfo{person}{Wenbo Wang}, \bibinfo{person}{Tien~N
  Nguyen}, \bibinfo{person}{Shaohua Wang}, \bibinfo{person}{Yi Li},
  \bibinfo{person}{Jiyuan Zhang}, {and} \bibinfo{person}{Aashish Yadavally}.}
  \bibinfo{year}{2023}\natexlab{}.
\newblock \showarticletitle{DeepVD: Toward Class-Separation Features for Neural
  Network Vulnerability Detection}. In \bibinfo{booktitle}{\emph{2023 IEEE/ACM
  45th International Conference on Software Engineering (ICSE)}}. IEEE,
  \bibinfo{pages}{2249--2261}.
\newblock


\bibitem[Wei et~al\mbox{.}(2022)]%
        {wei2022chain}
\bibfield{author}{\bibinfo{person}{Jason Wei}, \bibinfo{person}{Xuezhi Wang},
  \bibinfo{person}{Dale Schuurmans}, \bibinfo{person}{Maarten Bosma},
  \bibinfo{person}{Ed Chi}, \bibinfo{person}{Quoc Le}, {and}
  \bibinfo{person}{Denny Zhou}.} \bibinfo{year}{2022}\natexlab{}.
\newblock \showarticletitle{Chain of thought prompting elicits reasoning in
  large language models}.
\newblock \bibinfo{journal}{\emph{arXiv preprint arXiv:2201.11903}}
  (\bibinfo{year}{2022}).
\newblock


\bibitem[Wen et~al\mbox{.}(2023)]%
        {Wen2023vuldetect}
\bibfield{author}{\bibinfo{person}{Xin{-}Cheng Wen}, \bibinfo{person}{Yupan
  Chen}, \bibinfo{person}{Cuiyun Gao}, \bibinfo{person}{Hongyu Zhang},
  \bibinfo{person}{Jie~M. Zhang}, {and} \bibinfo{person}{Qing Liao}.}
  \bibinfo{year}{2023}\natexlab{}.
\newblock \showarticletitle{Vulnerability Detection with Graph Simplification
  and Enhanced Graph Representation Learning}. In
  \bibinfo{booktitle}{\emph{45th {IEEE/ACM} International Conference on
  Software Engineering, {ICSE} 2023, Melbourne, Australia, May 14-20, 2023}}.
  \bibinfo{publisher}{{IEEE}}, \bibinfo{pages}{2275--2286}.
\newblock
\urldef\tempurl%
\url{https://doi.org/10.1109/ICSE48619.2023.00191}
\showDOI{\tempurl}


\bibitem[Yamaguchi et~al\mbox{.}(2014)]%
        {yamaguchi2014modeling}
\bibfield{author}{\bibinfo{person}{Fabian Yamaguchi}, \bibinfo{person}{Nico
  Golde}, \bibinfo{person}{Daniel Arp}, {and} \bibinfo{person}{Konrad Rieck}.}
  \bibinfo{year}{2014}\natexlab{}.
\newblock \showarticletitle{Modeling and discovering vulnerabilities with code
  property graphs}. In \bibinfo{booktitle}{\emph{2014 IEEE Symposium on
  Security and Privacy}}. IEEE, \bibinfo{pages}{590--604}.
\newblock


\bibitem[Ying et~al\mbox{.}(2019)]%
        {ying2019gnnexplainer}
\bibfield{author}{\bibinfo{person}{Rex Ying}, \bibinfo{person}{Dylan
  Bourgeois}, \bibinfo{person}{Jiaxuan You}, \bibinfo{person}{Marinka Zitnik},
  {and} \bibinfo{person}{Jure Leskovec}.} \bibinfo{year}{2019}\natexlab{}.
\newblock \bibinfo{title}{GNNExplainer: Generating Explanations for Graph
  Neural Networks}.
\newblock
\newblock
\showeprint[arxiv]{1903.03894}~[cs.LG]


\bibitem[Zhang et~al\mbox{.}(2022)]%
        {zhang2022diet}
\bibfield{author}{\bibinfo{person}{Zhaowei Zhang}, \bibinfo{person}{Hongyu
  Zhang}, \bibinfo{person}{Beijun Shen}, {and} \bibinfo{person}{Xiaodong Gu}.}
  \bibinfo{year}{2022}\natexlab{}.
\newblock \showarticletitle{Diet code is healthy: Simplifying programs for
  pre-trained models of code}. In \bibinfo{booktitle}{\emph{Proceedings of the
  30th ACM Joint European Software Engineering Conference and Symposium on the
  Foundations of Software Engineering}}. \bibinfo{pages}{1073--1084}.
\newblock


\bibitem[Zhou et~al\mbox{.}(2020)]%
        {zhou2020graph}
\bibfield{author}{\bibinfo{person}{Jie Zhou}, \bibinfo{person}{Ganqu Cui},
  \bibinfo{person}{Shengding Hu}, \bibinfo{person}{Zhengyan Zhang},
  \bibinfo{person}{Cheng Yang}, \bibinfo{person}{Zhiyuan Liu},
  \bibinfo{person}{Lifeng Wang}, \bibinfo{person}{Changcheng Li}, {and}
  \bibinfo{person}{Maosong Sun}.} \bibinfo{year}{2020}\natexlab{}.
\newblock \showarticletitle{Graph neural networks: A review of methods and
  applications}.
\newblock \bibinfo{journal}{\emph{AI open}}  \bibinfo{volume}{1}
  (\bibinfo{year}{2020}), \bibinfo{pages}{57--81}.
\newblock


\bibitem[Zhou et~al\mbox{.}(2019)]%
        {zhou2019devign}
\bibfield{author}{\bibinfo{person}{Yaqin Zhou}, \bibinfo{person}{Shangqing
  Liu}, \bibinfo{person}{Jingkai Siow}, \bibinfo{person}{Xiaoning Du}, {and}
  \bibinfo{person}{Yang Liu}.} \bibinfo{year}{2019}\natexlab{}.
\newblock \showarticletitle{Devign: Effective vulnerability identification by
  learning comprehensive program semantics via graph neural networks}. In
  \bibinfo{booktitle}{\emph{In Proceedings of the 33rd International Conference
  on Neural Information Processing Systems}}. \bibinfo{pages}{10197–10207}.
\newblock


\end{thebibliography}
